\documentclass[twocolumn,secnumarabic,amssymb,
amsmath,nobibnotes, aps, showpacs,prl]{revtex4}

\usepackage{graphicx}
\usepackage{bm}

\def\be{\begin{equation}}
\def\ee{\end{equation}}

\begin{document}

\title{CMB Beam Systematics:\\
Impact on Lensing Parameter Estimation}
\ \\
\author{N.J. Miller, M. Shimon, B.G. Keating}
\affiliation{Center for Astrophysics and Space Sciences, University
of California, San Diego, 9500 Gilman Drive, La Jolla, CA, 92093-0424}
                                                                
\date{January 26}


\begin{abstract}

The cosmic microwave background (CMB) is a rich source of 
cosmological information. Thanks to the simplicity and linearity of the 
theory of cosmological perturbations, observations of the CMB's polarization and 
temperature anisotropy can reveal the parameters which describe the contents, 
structure, and evolution of the cosmos. 
Temperature anisotropy is necessary but not sufficient to fully mine 
the CMB of its cosmological information as it 
is plagued with various parameter degeneracies. 
Fortunately, CMB polarization breaks many of these 
degeneracies and adds new 
information and increased precision. 
Of particular interest is the CMB's B-mode polarization 
which provides a 
handle on several cosmological parameters 
most notably the tensor-to-scalar ratio, $r$, and is sensitive to 
parameters which govern the growth of large scale structure (LSS)  
and evolution of the gravitational potential. These imprint 
CMB temperature anisotropy and cause E-to-B-mode polarization 
conversion via gravitational lensing. However, 
both primordial gravitational-wave- and secondary lensing-induced  
B-mode signals are 
very weak and therefore prone to various foregrounds and systematics.
In this work we use Fisher-matrix-based estimations 
and apply, for the first time, Monte-Carlo Markov Chain (MCMC) 
simulations to determine the 
effect of beam systematics on the inferred cosmological parameters from 
five upcoming experiments: PLANCK, POLARBEAR, SPIDER, QUIET+CLOVER 
and CMBPOL. 
We consider beam systematics which couple the beam substructure to the 
gradient of temperature anisotropy and polarization 
(differential beamwidth, pointing offsets and ellipticity) 
and beam systematics due to differential beam normalization 
(differential gain) and orientation (beam rotation) 
of the polarization-sensitive axes (the latter two 
effects are insensitive to the beam substructure).
We determine allowable levels of beam systematics for given  
tolerances on the induced parameter errors and check for possible biases in the inferred 
parameters concomitant with potential increases in the statistical uncertainty.
All our results are scaled to the `worst case scenario'.
In this case, and for our tolerance levels the beam rotation should not exceed the few-degree 
to sub-degree level, typical ellipticity is required to be 1\%, 
the differential gain allowed level is few parts in $10^{3}$ to $10^{4}$, 
differential beam width upper limits are of the sub-percent level 
and differential pointing should not exceed the few- to sub-arc sec level.

\end{abstract}

\maketitle

\section{Introduction}

The standard cosmological model accounts for a multitude of phenomena occurring over 
orders of magnitude of length and angular scales throughout the entire history of cosmological evolution.
Remarkably, doing so only requires about a dozen parameters. 
Perhaps one of the most useful cosmological 
probes is cosmic microwave background (CMB) temperature anisotropy 
whose physics is well understood. 
Complementary cosmological probes can assist in breaking some 
of the degeneracies inherent in the CMB and further tighten the constraints on 
the inferred cosmological parameters. Temperature anisotropy alone cannot 
capture all the cosmological information in the CMB, and its polarization probes 
new directions in parameter space. 
B-mode polarization observations are noise-dominated but 
the robust secondary signal associated with gravitational lensing, 
which is known up to an uncertainty factor of two on all relevant scales, 
is at the threshold of detection by upcoming CMB experiments. 
The lensing signal may have been detected already through 
its signature on the CMB anisotropy as reported 
recently by ACBAR (Reichardt et al. [1]). 
Lensing by the large scale structure (LS) also converts primordial 
E-mode to secondary B-mode.  
When high fidelity B-mode data are available 
a wealth of information from the inflationary era 
(Zaldarriaga \& Seljak [2], Kamionkowski, Kosowsky \& Stebbins [3]), 
and cosmological parameters that control the evolution of small scale density perturbations 
(such as the running of the spectral index of primordial density perturbations, 
neutrino mass and dark energy equation of state), will be extracted from the CMB.
At best, B-mode polarization from lensing is a factor of three times smaller than 
the primordial E-mode polarization, 
so it is prone to contamination by both astrophysical 
foregrounds and instrumental systematics.
It is mandatory to account for, and remove when possible, all sources of spurious B-mode in 
analyzing upcoming CMB data, especially those generated by temperature leakage 
due to beam mismatch, since temperature anisotropy is several orders of magnitude larger than 
the expected B-mode level produced by lensing. 

Beam systematic have been discussed extensively 
(Hu, Hedman \& Zaldarriaga [4], Rosset et al. [5], O'Dea, 
Challinor \& Johnson [6], Shimon et al. [7]). 
All the effects are associated with beam imperfections or 
beam mismatch in dual beam experiments, i.e. where the polarization is obtained 
by differencing two signals which are measured simultaneously by two beams 
with two orthogonal polarization axes. Fortunately, several of these effects 
(e.g. differential 
gain, differential beam width and the first order pointing error - `dipole';
Hu, Hedman \& Zaldarriaga [4], O'Dea, Challinor \& Johnson [6], Shimon et al. [7]) 
are reducible with an ideal scanning strategy 
and otherwise can be cleaned from the data set by virtue of their 
non-quadrupole nature which distinguishes them from genuine CMB polarization 
signals. 
Other spurious polarization signals, such as 
those due to differential ellipticity of the beam, second order 
pointing errors and differential rotation, persist even in the 
case of ideal scanning strategy and perfectly mimic CMB 
polarization. These represent the minimal 
spurious B-mode signal, residuals which will plague every 
polarization experiment. We refer to them in the following as 
`irreducible beam systematics'. We assume throughout that beam parameters 
are spatially constant. 
Two recent works (Kamionkowki [8] and Su, Yadav \& Zaldarriaga [9]) considered 
the effect of spatially-dependent systematic beam-rotation and differential 
gain, respectively. This scale-dependence and the associated new angular scale 
induce non-trivial higher order correlation functions through non-gaussianities 
which can be both used to optimally remove the space-dependent component of 
beam rotation [8] and mimic the CMB lensing signal, thereby biasing the 
quadratic estimator of the lensing potential [9].

To calculate the effect of beam systematics we invoke the Fisher 
information-matrix formalism as well as 
Monte Carlo simulations of parameter extraction, the latter 
for the first time. 
Our objective is to determine the susceptibility of the 
above mentioned, and other, cosmological parameters to beam systematics.
For the Fisher-matrix-based method and the Monte Carlo simulations 
we calculate the underlying power spectrum using CAMB 
(Lewis, Challinor \& Lasenby [10]). The Monte-Carlo 
simulations are carried out with COSMOMC (Lewis \& Bridle [11]). We represent the extra noise 
due to beam systematics by analytic approximations (Shimon et al. [7]) 
and include lensing extraction in the parameter inference process, 
following Kaplinghat, Knox \& Song [12] and Lesgourgues et al. [13] 
(see also Perotto et al. [14] for the Monte Carlo simulations) 
for neutrino mass (and other cosmological parameters) reconstruction from CMB data.

This paper illustrates the effect of beam systematics and its 
propagation to parameter 
estimation and error forecasts for upcoming experiments.
Our main concern is the effect on the following cosmological parameters: 
the tensor-to-scalar ratio $r$,
the total neutrino mass $M_{\nu}$ (assuming three degenerate species), 
tilt of the scalar index $\alpha$, 
dark energy equation of state $w$, and the 
spatial curvature, $\Omega_{k}$. 
The lensing-induced B-mode signal is sensitive to all parameters (except the tensor-to-scalar ratio) 
and peaks at few arcminute scales, while the 
tensor-to-scalar ratio depends on the energy scale of inflation 
and the primordial signal peaks at the characteristic horizon size at last scattering, 
$\approx 2^{\circ}$. We note that while the LSS-induced and primordial tensor power 
B-mode spectra are sub-$\mu K$ the {\it shape} of the primordial B-mode spectrum is known
(only its amplitude is unknown, Keating [15]) and the secondary LSS-induced 
B-mode is guaranteed to exist by virtue of 
the known existence of LSS and E-mode polarization.

The paper is organized as follows. 
We describe the formalism of beam systematics for general non-gaussian beams 
and provide a cursory description of a critical tool to mitigate polarization systematics 
-a half wave plate (HWP), in section 2.
The effect of lensing on parameter extraction within the standard quadratic-estimators 
formalism is discussed in section 3.
The essentials of the Fisher matrix formalism are given in section 4 as 
well as some details on the Monte Carlo simulations invoked here. 
Our results are described in section 5 and we conclude with a discussion of 
our main findings in section 6.   

\section{Beam systematics}

Beam systematics due to optical imperfections depend on both the underlying sky, 
the properties of the polarimeter and on the scanning strategy. 
Temperature anisotropy leaks to polarization when the output of two 
slightly different beams with orthogonal polarization-sensitive directions is being differenced. 
A trivial example is the effect of differential gain. 
If the two beams have the same shape, width, etc. except for different overall response, 
i.e. normalization, the difference of the measured intensity will result in a non-vanishing 
polarization signal. Similarly, if two circular beams slightly differ by their width 
this will again induce a non-vanishing polarization upon taking the difference 
(see Fig. 2 at Shimon et al. [7]). 
The spurious polarization will be proportional to temperature fluctuations on scales comparable to 
the difference in beamwidths, which, due to the circular symmetry of the problem, will be proportional 
to second order gradients of the temperature anisotropy. To eliminate these effects
this beam imperfection has to couple to non-ideal scanning strategy as described in Shimon et al. 
[7] and below. A closely related effect, which does not couple to scanning strategy, is the 
effect of differential beam ellipticity. Here, the spurious polarization scales as the second order 
gradient of the temperature anisotropy to leading order. Another effect, widely described in the literature, 
which again couples beam asymmetry and temperature anisotropy to scanning strategy is the effect due 
to differential pointing. The idea is simple; if two beams point at two slightly different directions 
they will statistically measure two different intensities proportional to fluctuations of the background 
radiation on these particular scales. The difference, which may be naively regarded as polarization, 
is non-vanishing in this case provided the scanning strategy is non-ideal and contains either a 
dipole and/or an octupole (Shimon et al. [7]). 
Finally, the effect of beam rotation we consider in this work is due to 
uncertainty in the overall beam orientation. This mixes the Q and U Stokes parameters and as a 
result also leaks E to B and vice versa.
A constructive order-of-magnitude example is the effect of differential pointing. This effect depends on 
the temperature gradient to first order. The rms CMB temperature gradients at the $1^{\circ}$, $30'$, 
$10'$, $5'$ and $1'$ scales are $\approx$ 1.4, 1.5, 3.5, 2.5 and 0.2 $\mu K/{\rm arcmin}$, respectively.
Therefore, any temperature difference measured with a dual-beam experiment 
(with typical beamwidth few arcminutes) with 
a $\approx 1'$ pointing error will result in a $\approx 1\mu$K systematic 
polarization which has the potential to overwhelm the B-mode signals.
 
Similarly, the systematic induced by differential ellipticity results from the variation 
of the underlying temperature anisotropy along the two polarization-sensitive directions 
which, in general, differ in scale depending on the mean beamwidth, degree of ellipticity 
and the tilt of the polarization-sensitive direction with respect to the ellipse's 
principal axes. For example, the temperature difference measured along the major and minor axes 
of a $1^{\circ}$ beam with a 2\% ellipticity scales as the second gradient of the 
underlying temperature which on this scale is $\approx 0.2\mu K/{\rm arcmin^{2}}$ 
and the associated induced polarization is therefore expected to be on the $\approx\mu K$ 
level.   
The spurious signals due to pointing error, 
differential beamwidth and beam ellipticity all peak at angular scales comparable to the beam size. If the beam 
size is $\approx 1^{\circ}$ the beam systematics mainly affect the deduced tensor-to-scalar ratio, $r$.
If the polarimeter's beamwidth is a few arcminutes the associated systematics will impact the measured 
neutrino mass $m_{\nu}$, spatial curvature $\Omega_{k}$, running of the scalar spectral index 
$\alpha$ and the dark energy equation of state $w$ (which strongly affects the lensing-induced B-mode signal). 
It can certainly be the case that other cosmological parameters will be affected as well.

Two other spurious polarization signals we explore are 
due to differential gain and differential rotation; 
these effects are associated with different beam `normalizations' and orientation, respectively, and 
are independent of the coupling between beam substructure and the underlying temperature perturbations.
In particular, they have the same scale dependence as the primordial 
temperature anisotropy and polarization power spectra 
(as long as the differential gain and beam rotation are spatially independent; this of course changes if they depend on space [8], [9]), respectively, and their peak impact 
will be on scales associated with the CMB's temperature anisotropy ($\approx 1^{\circ}$) 
and polarization ($\approx 10'$).

\subsection{Mathematical Formalism}

We work entirely in Fourier space and in this section we generalize our results 
(Shimon et al. [7]) to the case of the most general beam shapes.
Although the tolerance levels on the beam parameters we derive in sections 4 and 5 
are based on the assumption of elliptical beams, they can be easily generalized 
to arbitrary beam shape, given the beam profile, as we describe below.
This can be used to adapt our results to actual measured beam maps incorporating 
other classes of beam non-ideality such as sidelobes.

We expand the temperature anisotropy and $Q$ and $U$ Stokes parameters in 2-D
plane waves since for sub-beam scales this is a good approximation. 
While the (spin-0) temperature anisotropy is expanded in scalar plane waves $e^{i{\bf l}\cdot{\bf r}}$, 
the (spin $\pm$2) polarization tensor $Q+iU$ is expanded in tensor plane waves
$e^{i{\bf l}\cdot{\bf r}}e^{\pm 2i(\phi_{{\bf l}}-\phi_{{\bf r}})}$ where 
$\phi_{{\bf r}}$ is the angle defining the direction of the radius-vector r 
in real space as conventional (in an arbitrarily coordinate system on the 
sky $\phi_{r}$ is the azimuthal angle along the line of sight) 
and $\phi_{{\bf l}}$ defines the direction of the 
wave-vector ${\bf l}$ in l-space in a coordinate-system fixed to the beam 
as defined below, in Eq. (2).
Since in real space the temperature and polarization fields are convolved
with the polarimeter's beams,
these expressions are simply the product of their Fourier transforms in 
Fourier space. 
For a general beam $B({\bf r})$ the measured 2-D polarized beam map may exhibit 
a pointing error ${\bf \rho}$. In this case, the Fourier transform of the beam function 
acquires a phase
\begin{eqnarray}
\tilde{B}({\bf l})\rightarrow\tilde{B}({\bf l})\exp(i{\bf l}\cdot{\bf\rho}).
\end{eqnarray}
It is useful to switch to polar coordinates at this point, where we define
\begin{eqnarray}
l_{x}&=&l\cos(\phi_{{\bf l}}+\psi-\alpha)\nonumber\\
l_{y}&=&l\sin(\phi_{{\bf l}}+\psi-\alpha)\nonumber\\
\rho_{x}&=&\rho\cos\theta\nonumber\\
\rho_{y}&=&\rho\sin\theta
\end{eqnarray}
and $\alpha\equiv\beta+\theta+\psi$ is the angle 
of the polarization axis in a coordinate system 
fixed to the sky (Fig. 1 of Shimon et al. [7]). 
The Fourier representation of an {\it arbitrary} beam then becomes
\begin{eqnarray}
\tilde{B}({\bf l})&=&\int B({\bf r})e^{i{\bf l}\cdot{\bf r}}d^{2}{\bf r}\nonumber\\
&=&\int B({\bf r})e^{ilr\cos(\phi_{l}+\psi-\alpha-\phi_{r})
+il\rho\cos(\phi_{l}+\theta+\psi-\alpha-\phi_{r})}d^{2}{\bf r}\nonumber\\
&\equiv &\sum_{m,n}B_{m,n}(l)e^{i(m+n)(\phi_{l}-\alpha)} 
\end{eqnarray}
where
\begin{eqnarray}
B_{m,n}(l)&=&i^{m+n}J_{n}(l\rho)e^{i(m+n)\psi+in\theta}\nonumber\\
&\times &\int r dr J_{m}(lr)\int d\phi_{r}B({\bf r})e^{-i(m+n)\phi_{r}},
\end{eqnarray}
and in the last step we employed the expansion of 2-D plane waves in 
terms of cylindrical Bessel functions
\begin{eqnarray}
e^{il\rho\cos (\phi_{l}-\phi_{\rho})}=\sum_{n=-\infty}^{n=\infty}
i^{n}J_{n}(l\rho)e^{in(\phi_{l}-\phi_{\rho})}.
\end{eqnarray}
As in Shimon et al. [7], the optimal map constructed from the CMB data depends on the measurements as
\begin{eqnarray}
\tilde{m}({\bf p})&=&\left(\sum_{t,j\in {\bf p}} A_{j}^{T}({\bf p},t)A_{j}({\bf p},t)\right)^{-1}\nonumber\\
&\times &\left(\sum_{t,j\in {\bf p}}A_{j}^{T}({\bf p},t)d_{j}({\bf p},t)\right)
\end{eqnarray}
where the sums run over all measurements of the pixel ${\bf p}$.  
The pointing vector $A$ is given by
\begin{eqnarray}
A=\left(1,\frac{1}{2}e^{2i\alpha({\bf p},t)},\frac{1}{2}e^{-2i\alpha({\bf p},t)}\right), 
\end{eqnarray}
$\alpha$ is a function of both the pixel {\bf p} and $t$, and $A^{T}$ is $A$ transposed.
Once the leading beam coefficients $B_{m,n}(l)$ have been calculated, the 
induced power spectra of the systematics can be calculated according to Eqs. (24), (33), (A.1) and (A.2) 
of Shimon et al. [7].

Several of the beam systematics can be mitigated by employing a rotating half wave plate (HWP) polarization modulator (e.g. Hanany et al. [16], Johnson et al. [17], MacTavish et al. [18]). These can operate in continuous or stepped rotation. 
When HWP modulators are included 
we replace the above scanning angle $\alpha({\bf p},t)$ with $\alpha({\bf p},t)+2\varphi t$ where $\varphi$ is the 
angular velocity of the HWP (O'Dea, Challinor \& Johnson [6]). Our deduced tolerance levels given below are 
presented in a fashion independent of the details of the scanning strategy; all the information about the scanning strategy is encapsulated in the functions $f_{1}$, $f_{2}$ and $f_{3}$:
\begin{eqnarray}
f_{1}&\equiv&\frac{1}{2}|\tilde{h}_{+}(-1,0)|^{2}\nonumber\\
f_{2}&\equiv&\frac{1}{2}|\tilde{h}_{+}(-1,-1)|^{2}+\frac{1}{2}|\tilde{h}_{+}(-1,1)|^{2}\nonumber\\
f_{3}&\equiv&\frac{1}{2}\langle\tilde{f}(0,1)\tilde{h}_{-}^{*}(1,-1)\rangle
\end{eqnarray}
where
\begin{eqnarray}
f(m,n)&\equiv&\langle e^{-i(2m+n)\alpha({\bf p},t)}\rangle\nonumber\\
h_{\pm}(m,n)&\equiv&\frac{1}{D}[f(m,n)-f(m\pm 2,n)\langle e^{\pm 4i\alpha({\bf p},t)}\rangle]\nonumber\\
D&\equiv& 1-\langle e^{-i4\alpha({\bf p},t)}\rangle\langle e^{i4\alpha({\bf p},t)}\rangle
\end{eqnarray}
and the angular brackets in $\langle e^{in\alpha({\bf p},t)}\rangle$
represent average over measurements of a single pixel ${\bf p}$, averaged over time.
In these averages $\alpha({\bf p},t)\rightarrow\alpha({\bf p},t)+2\varphi t$, and therefore even if the scanning strategy does not uniformly cover all polarization angles $\alpha$ of a given spatial pixel, the HWP mitigates the spurious polarization caused by beam systematic effects if integrated over long time intervals.
If the hexadecapole of the scanning strategy is negligible, the scanning strategy function $f_{1}$ depends only 
on the quadrupole moment of the scanning strategy while $f_{2}$ encapsulates information on both the dipole and octupole moments of the scanning strategy.

At this point, it is instructive to show how, in the case of an ideal scanning strategy the first order pointing effect vanishes. As can be seen from Table II this effect involves a convolution of the beam function and underlying temperature anisotropy power spectrum with $f_{2}$ in multipole space. From the above definitions 
\begin{eqnarray}
h_{+}(-1,-1)&=&1/D\left[\langle e^{3i\alpha}\rangle-\langle e^{-i\alpha}\rangle\langle e^{4i\alpha}\rangle\right]\nonumber\\
h_{+}(-1,1)&=&1/D\left[\langle e^{i\alpha}\rangle-\langle e^{-3i\alpha}\rangle\langle e^{4i\alpha}\rangle\right]\nonumber\\
h_{-}(1,-1)&=&h_{+}^{*}(-1,1)
\end{eqnarray}
and therefore if with each scanning angle $\alpha$ there is associated an angle $\alpha+180^{\circ}$ the $f_{2}$ (Eqs. 8 and 10) vanishes in real space and so does its Fourier transform. Note that even if the scanning strategy is {\it non-ideal} $f_{2}$ will vanish provided that for each angle $\alpha$ the angle $\alpha+180^{\circ}$ is sampled the same number of times per pixel. This suggests that the {\it dipole} systematic can be completely removed by removing all data points that contribute to $h_{+}(-1,-1)$ and $h_{+}(-1,1)$, i.e. those measurements at $\alpha$ for which $\alpha+180^{\circ}$ is not sampled. Similar considerations apply to $f_{1}$ which controls the level of 
the differential beamwidth- and differential gain-induced systematics (see Table II). 

\subsection{Simplifying Scan Strategy Effects}

When the polarization angle at each pixel on the sky is 
uniformly sampled the average $\langle e^{in\alpha}\rangle$ vanishes for every $n\neq 0$. 
In this case the scanning strategy is referred to as an {\it ideal scanning strategy}.
For uniform, but non-ideal, scanning strategies, the scanning 
functions $f_{1}$, $f_{2}$ and $f_{3}$ mentioned above 
(which are combinations of $\langle e^{in\alpha}\rangle$) 
are non-vanishing even when $n>0$ but uniform in real space. 
As a result their Fourier transforms are unnormalized delta-functions (the actual amplitudes 
are directly related to the average values $\langle e^{in\alpha}\rangle$), 
and the convolutions in Fourier space 
shown in Tables III-IV of Shimon et al. [7] become trivial. To determine the tolerance 
level for beam parameters we assume such uniform scanning strategies. 

A uniform scanning strategy is a particularly useful example.
A nearly-uniform scanning strategy 
can be reasonably approximated by a sum over a few lowest multipoles, such as
\begin{eqnarray}
\tilde{f}_{i}(l)=\sum_{l'=0}^{l'_{max}}f'_{i}(l')\delta(l-l')/(l-l').
\end{eqnarray}
Here $|l_{max}|$ is assumed sufficiently small, and $i=1$ or $2$, where $f_{1}$, $f_{2}$ 
and $f_{3}$ are defined in Eq. 8). In this case the $\ell$-mode mixing due 
to the convolution of the underlying power spectra and the scanning functions as in Table II 
($\star$ stands for convolution in multipole space). $\tilde{f}_{i}$ can be written as
\begin{eqnarray}
\tilde{f}_{i}\star C_{l}^{T}\approx\frac{f_{i}}{2\pi}\cdot C_{l}^{T} \nonumber \\
f_{i}\equiv\sum_{l'=1}^{l'_{max}}f'_{i}(l').
\end{eqnarray}
We assume here that 
the nonvanishing multipoles of $\tilde{f}_{i}$ are concentrated near 0, i.e. that the 
scanning strategy is non-ideal, yet approximately uniform. 
We have employed this simplifying assumption throughout.

\section{The Effect of Systematics on Lensing Reconstruction}

Gravitational lensing of the CMB is both a nuisance and a valuable cosmological tool 
(e.g. Zaldarriaga \& Seljak [19]). 
It certainly has the potential to complicate CMB data analysis due to the non-gaussianity 
it induces. However, it is also a unique probe of the growth of structure in the 
linear, and mildly non-linear, regimes (redshift of a few). 
Kaplinghat, Knox \& Song [12], Lesgourgues et al. [13], as well as others, have shown that with a {\it nearly ideal 
CMB experiment} (in the sense that instrumental noise as well as astrophysical foregrounds are negligibly small), neutrino mass limits can be improved by a factor of approximately four
by including lensing extraction in the data analysis using CMB data alone.
This lensing extraction process is not perfect; a fundamental residual noise will 
afflict any experiment, even ideal ones.
This noise will, in principle, propagate to the inferred cosmological parameters if 
the latter significantly depend on lensing extraction, e.g. neutrino mass, $\alpha$ and $w$.
It is important to illustrate first the effect of beam systematics on lensing reconstruction.
By optimally filtering the temperature and polarization Hu \& Okamoto [20] reconstructed 
the lensing potential from quadratic estimators.
It was shown that for experiments with ten times higher sensitivity than 
Planck, the EB estimator yields the tightest limits on the lensing potential. This 
result assumes no beam systematics which might significantly contaminate the observed 
B-mode. 

We illustrate the effect of differential beam rotation, ellipticity and differential pointing
(see Shimon et al. [7]) on the noise of lensing reconstruction 
with POLARBEAR (1200 detectors), CMBPOL-A 
(one of two toy experiments we consider for CMBPOL; 0.22$\mu K$ sensitivity and 5' beam) 
and a toy-model considered earlier by O'Dea, Challinor \& Johnson [6] which we refer to as QUIET+CLOVER 
in Figures 1, 2 and 3, respectively. 
These are perhaps the most pernicious systematics. Beam rotation  
induces cross-polarization which leaks the much larger 
E-mode to B-mode polarization and differential ellipticity leaks T to B. Both leak to B-mode 
in a way indistinguishable under rotation from a true B-mode signal.
The rotation and ellipticity 
parameters ($\varepsilon$ and $e$, respectively) we considered range from 0.01 to 0.20 
($e$ is dimensionless and $\varepsilon$ is given in radians). 
The differential pointing $\rho$, was set to 1\% and 10\% of the beamwidth while the dipole and octupole components of the scanning strategy were set to the `worst case scenario' $f_{2}=2\pi$, i.e. the unlikely situation where all `hits' at a given pixel take place at the same polarization angle $\alpha$ (again, for ideal scanning strategy $f_{2}=0$ and the {\it dipole} effect due to differential pointing vanishes). 
Note for POLARBEAR (Figure 1) with $\varepsilon,e=0.2$ the lensing potential can be reconstructed up to $l\approx 200$, while with no beam rotation it can be reconstructed up to $l\approx 250$.
However, with CMBPOL-A (Figure 2) lensing reconstruction degrades significantly 
in the presence of beam rotation (from good reconstruction up to l$\approx$600 in 
the systematics-free case down to l$\approx$ 250 when $\varepsilon,e=0.2$ 
and $\rho=0.5'$ (in case $f_{2}=2\pi$)).
The reason for the qualitative difference is that for experiments with sensitivities 
comparable to PLANCK or POLARBEAR, the best estimator of the lensing potential comes from the TT, TE
and EE correlations (depending on scale $l$) and the cross-correlations involving B-mode are only secondary in probative power
(see top left panel of Fig.1). 
Therefore, lensing reconstruction 
for these experiments is hardly affected by beam systematics (we ignored the negligible 
beam systematics' effect on temperature anisotropy and considered only those of E and B). 
In contrast, as can be 
seen from Figure 2, CMBPOL-A's lensing reconstruction is significantly degraded since 
its lensing reconstruction is dominated by the contribution of the EB estimator 
for all relevant multipoles (see top left panel of Fig.2).  
The modified noise in reconstructing the lensing potential, $N_{l}^{dd}$, is 
consistently substituted into our Fisher matrix and Monte Carlo simulations 
(below, we summarize the relevant expressions of the quadratic estimators method).
 
Following Hu \& Okamoto [20] the MV noise on the lensing deflection 
angle reconstructed power spectrum $C_{l}^{dd}$ is
\begin{eqnarray}
N_{MV}^{dd}=\left[\sum_{\alpha\beta}(N^{-1})_{\alpha\beta}\right]^{-1}
\end{eqnarray}
where 
\begin{eqnarray}
N_{\alpha\beta}(L)&=&L^{-2}A_{\alpha}(L)A_{\beta}(L)\int\frac{d^{2}l_{1}}{(2\pi)^{2}}
F_{\alpha}(\bf{l}_{1},\bf{l}_{2})\left(F_{\beta}(\bf{l}_{1},\bf{l}_{2})\right.\nonumber\\
&\times &\left.C_{l_{1}}^{x_{\alpha},x_{\beta}}C_{l_{2}}^{x'_{\alpha},x'_{\beta}}
+F_{\beta}(\bf{l}_{2},\bf{l}_{1})
C_{l_{1}}^{x_{\alpha},x'_{\beta}}C_{l_{2}}^{x'_{\alpha},x_{\beta}}\right)\nonumber\\
A_{\alpha}(L)&\equiv &L^{2}\left[\int\frac{d^{2}{\bf l_{1}}}{(2\pi)^{2}}h_{\alpha}({\bf l_{1}},{\bf l_{1}})F_{\alpha}({\bf l_{1}},{\bf l_{1}})\right]^{-1}
\end{eqnarray}
and $\alpha$ stands for one of the pairings TT,TE,EE,TB and EB (BB does not participate in these combinations). The coupling
takes place between different modes $l_{1}$ and $l_{2}$. When $\alpha=TT$ or $EE$
\begin{eqnarray}
F_{\alpha}({\bf l}_{1},{\bf l}_{2})\rightarrow
\frac{h_{\alpha}({\bf l}_{1},{\bf l}_{2})}{2C'^{xx}_{l_{1}}C'^{xx}_{l_{2}}}, 
\end{eqnarray}
and when $\alpha=TB$ or $EB$
\begin{eqnarray}
F_{\alpha}({\bf l}_{1},{\bf l}_{2})\rightarrow\frac{h_{\alpha}({\bf l}_{1},{\bf l}_{2})}{C'^{xx}_{l_{1}}C'^{x'x'}_{l_{2}}} 
\end{eqnarray}
where $C'_{l}$ are the {\it observed} power spectra, i.e. including lensing, main-beam 
dilution on small scales, and in principle - {\it beam systematics} (see Tables I and II). 
The latter mainly affect the B-mode polarization, and as a result, the EB estimator. 
A list of $h_{\alpha}({\bf l}_{1},{\bf l}_{2})$ can be found in Hu \& Okamoto [20].
We have used the publically available code [21] 
employed in Lesgourgues et al. [13] and in Perotto et al. [14]. The code is based on the formalism developed in Okamoto \& Hu [22], an extension of Hu \& Okamoto [20] to the full-sky, to calculate the noise level in lensing reconstruction.

\begin{figure*}
\begin{tabular}{cc}
\includegraphics[width=\columnwidth]{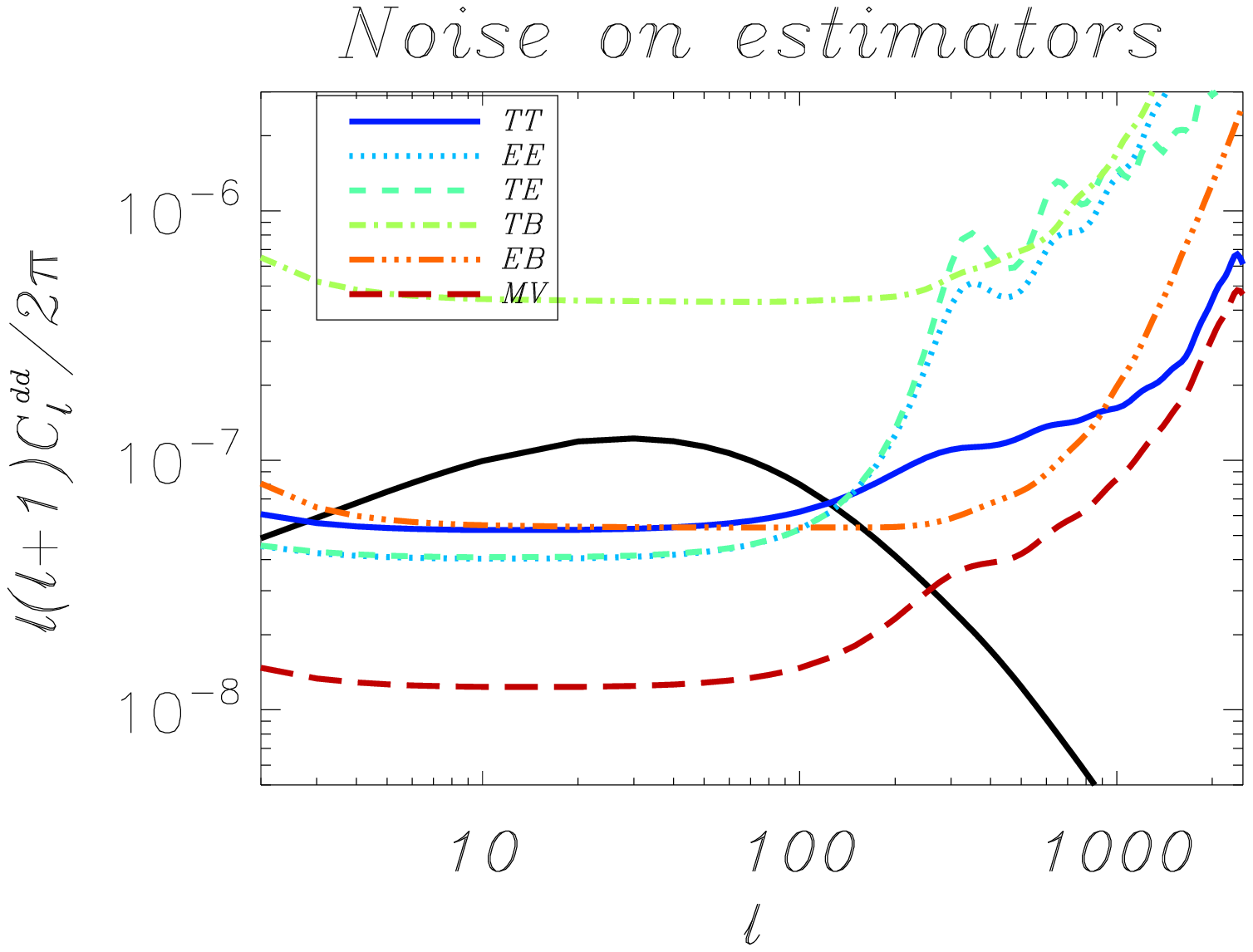} &
\includegraphics[width=\columnwidth]{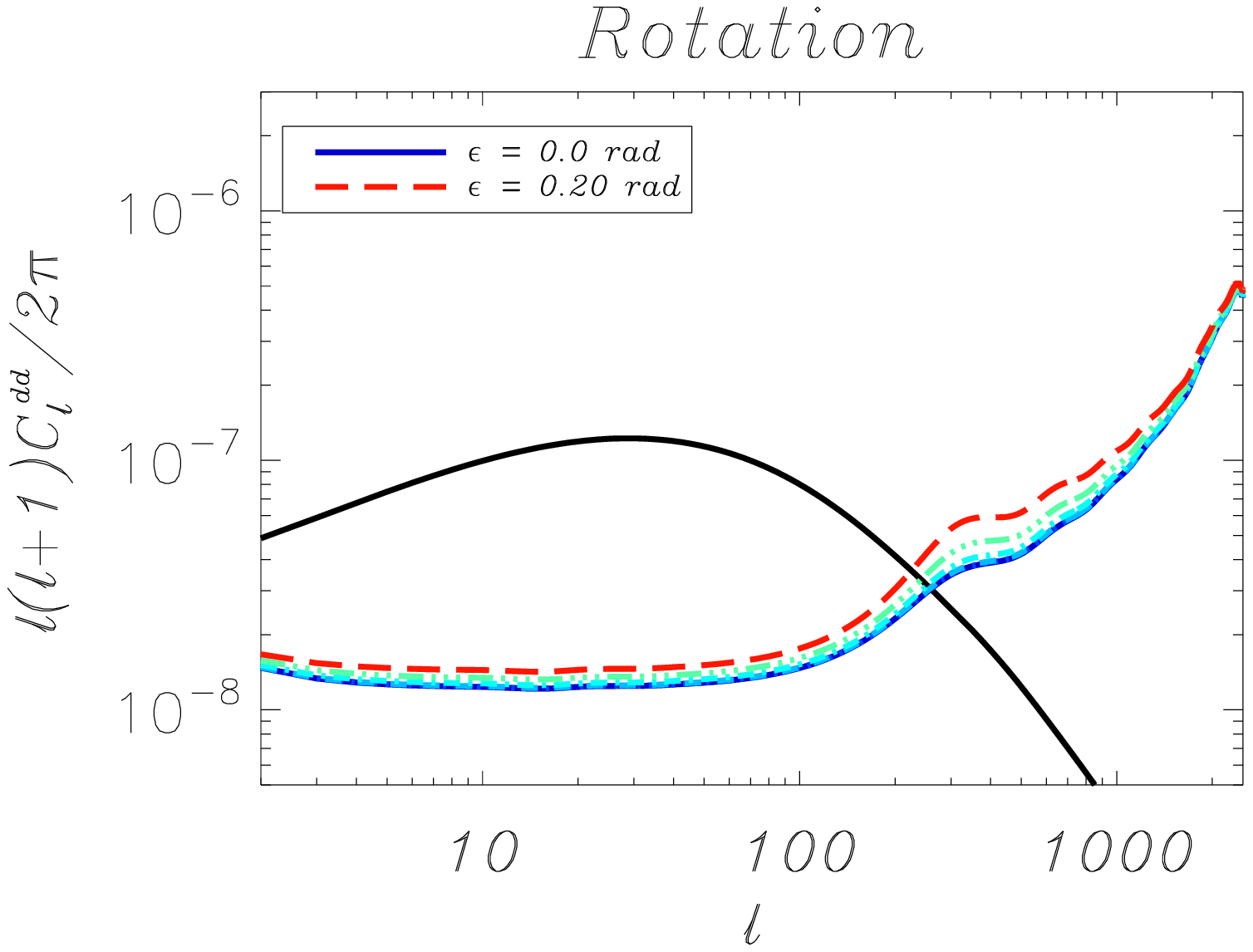} \\
\includegraphics[width=\columnwidth]{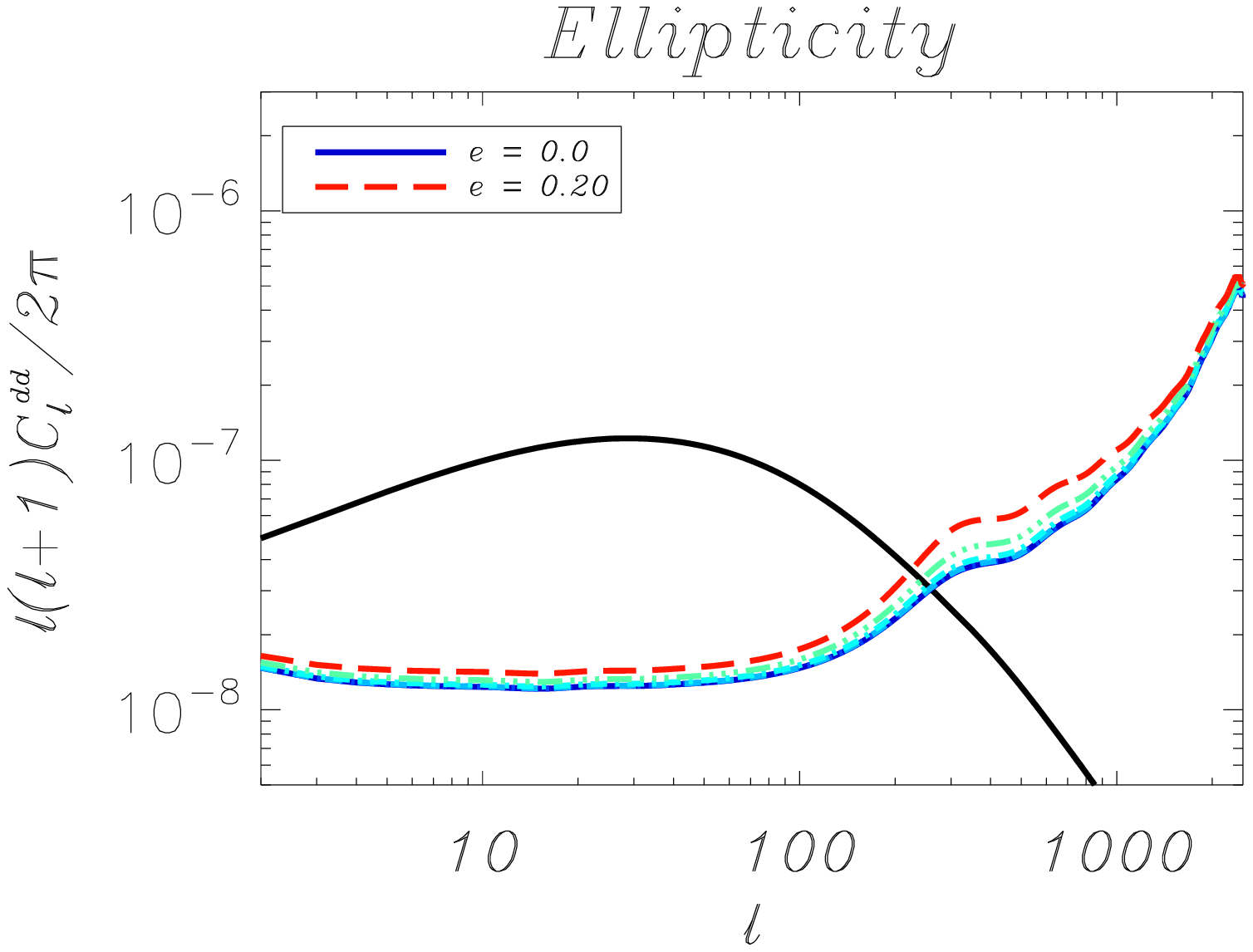} &
\includegraphics[width=\columnwidth]{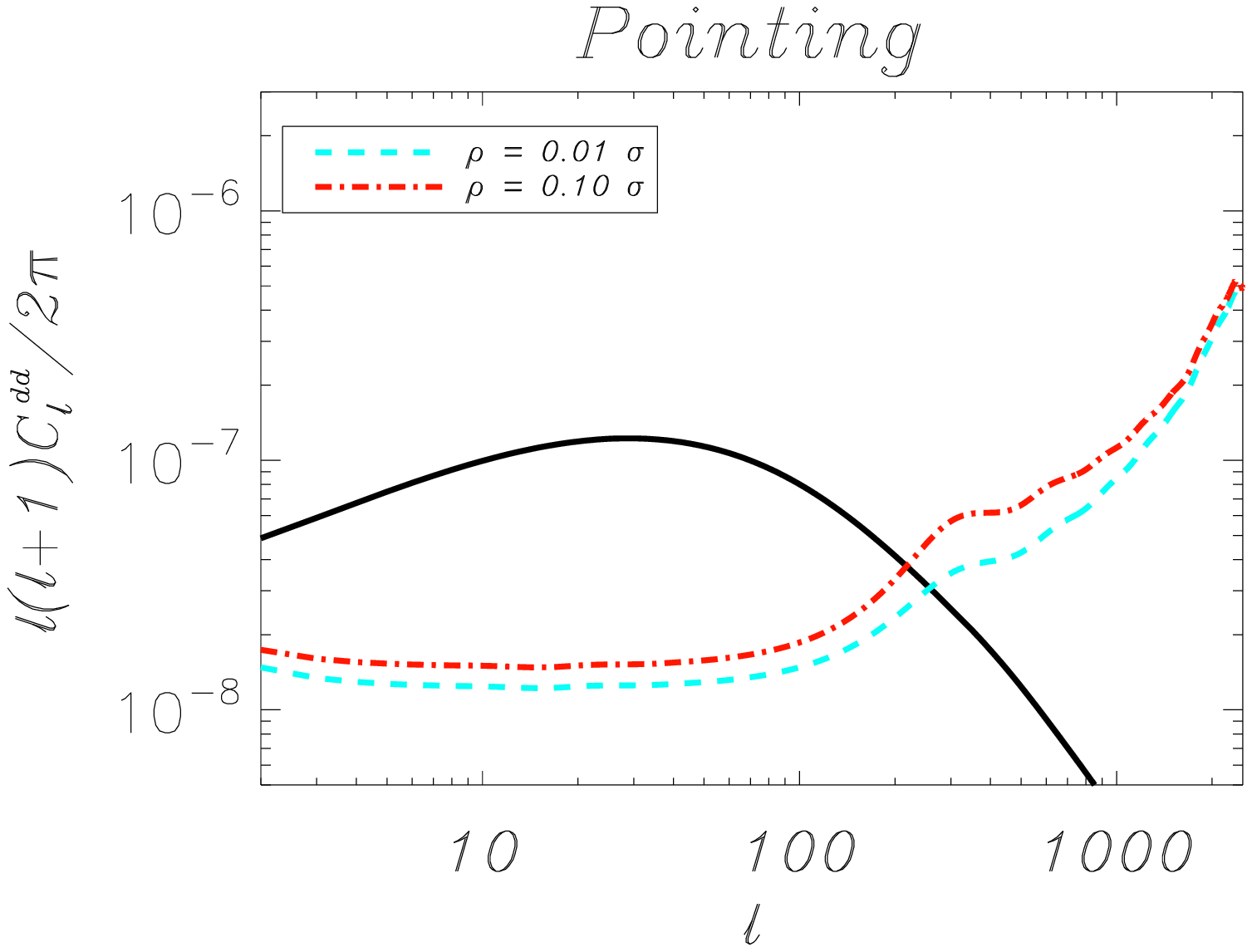}
\end{tabular}
\caption{For all panels the solid black curve is the deflection angle power spectrum $C_{l}^{dd}$ 
caused by gravitational lensing by the LSS.
{\it Top left}: The noise (with no systematics) in lensing-reconstruction from the quadratic optimal filters for POLARBEAR 
TT (solid dark blue), EE (dot light blue), TE (dashed green), TB (dot-dash yellow), EB (double-dot-dash orange) and MV (dashed dark red). 
For POLARBEAR sensitivity and angular resolution, the lowest-noise estimator is one of the EE and TE estimators depending on the angular size. 
{\it Top right}: Noise in lensing reconstruction for POLARBEAR with the MV estimator 
including the effects of the most pernicious irreducible cross-polarization systematic: 
differential rotation. 
Differential rotation values are (bottom to top) 
$\varepsilon=$0.01, 0.02, 0.05, 0.10 and 0.20 radian, respectively. 
High signal-to-noise deflection angle reconstruction can be obtained over 
nearly a decade of angular scale. The lensing reconstruction is not significantly affected by systematics because of the significant 
contribution of the temperature to the MV estimator.
{\it Bottom left}: Noise in lensing reconstruction for POLARBEAR with the MV estimator 
including the most pernicious {\it irreducible} instrumental polarization systematics: 
differential ellipticity. 
Differential ellipticity values are (bottom to top) 
$e=$0.01, 0.02, 0.05, 0.10 and 0.20, respectively (we assume $\psi=45^{\circ}$). The same explanation for insensitivity to differential 
rotation applies here for differential ellipticity $e$; 
the best estimator for this experiment is derived from temperature correlations which are 
hardly affected by beam systematics (and completely ignored in this analysis).
{\it Bottom right}: Noise in lensing reconstruction for POLARBEAR with the MV estimator 
for the most pernicious {\it reducible} instrumental polarization systematics: differential pointing
with 1\% and 10\% pointing errors (i.e. $\rho=0.01\sigma$ and $\rho=0.1\sigma$, respectively) 
under the `worst case' assumption that the scanning-strategy-related function $f_{2}=2\pi$.} 
\end{figure*}  

\begin{figure*}
\begin{tabular}{cc}
\includegraphics[width=\columnwidth]{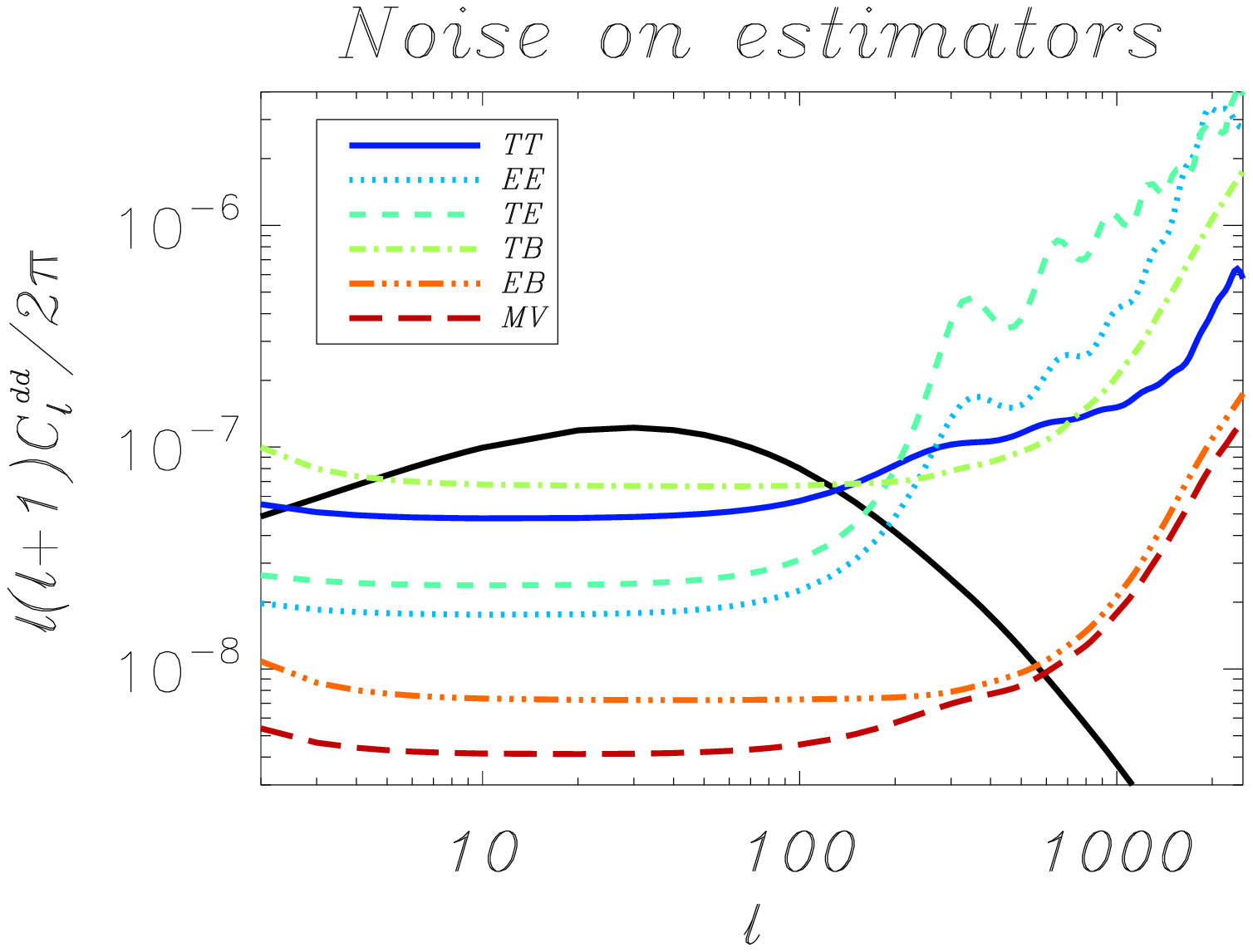} &
\includegraphics[width=\columnwidth]{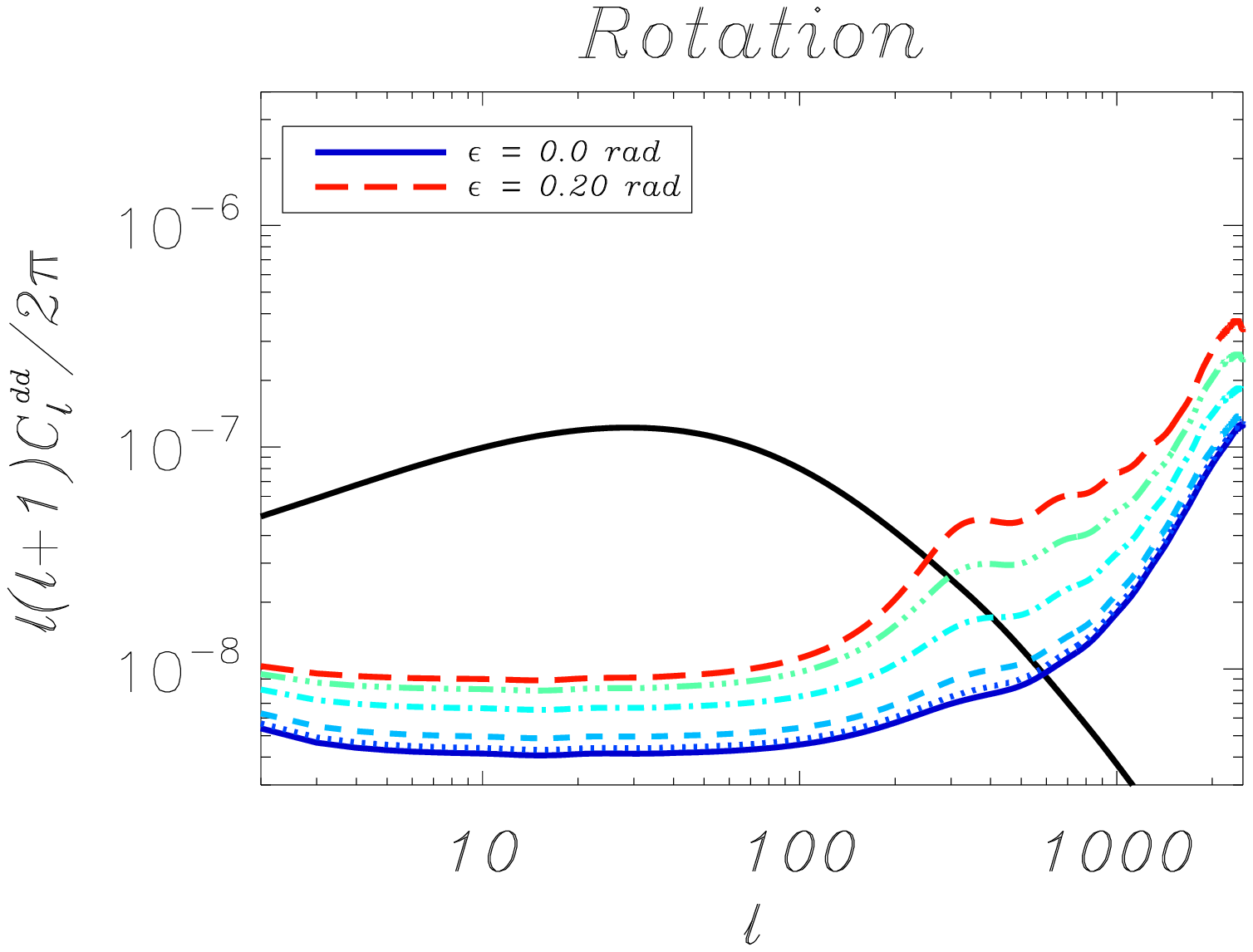} \\
\includegraphics[width=\columnwidth]{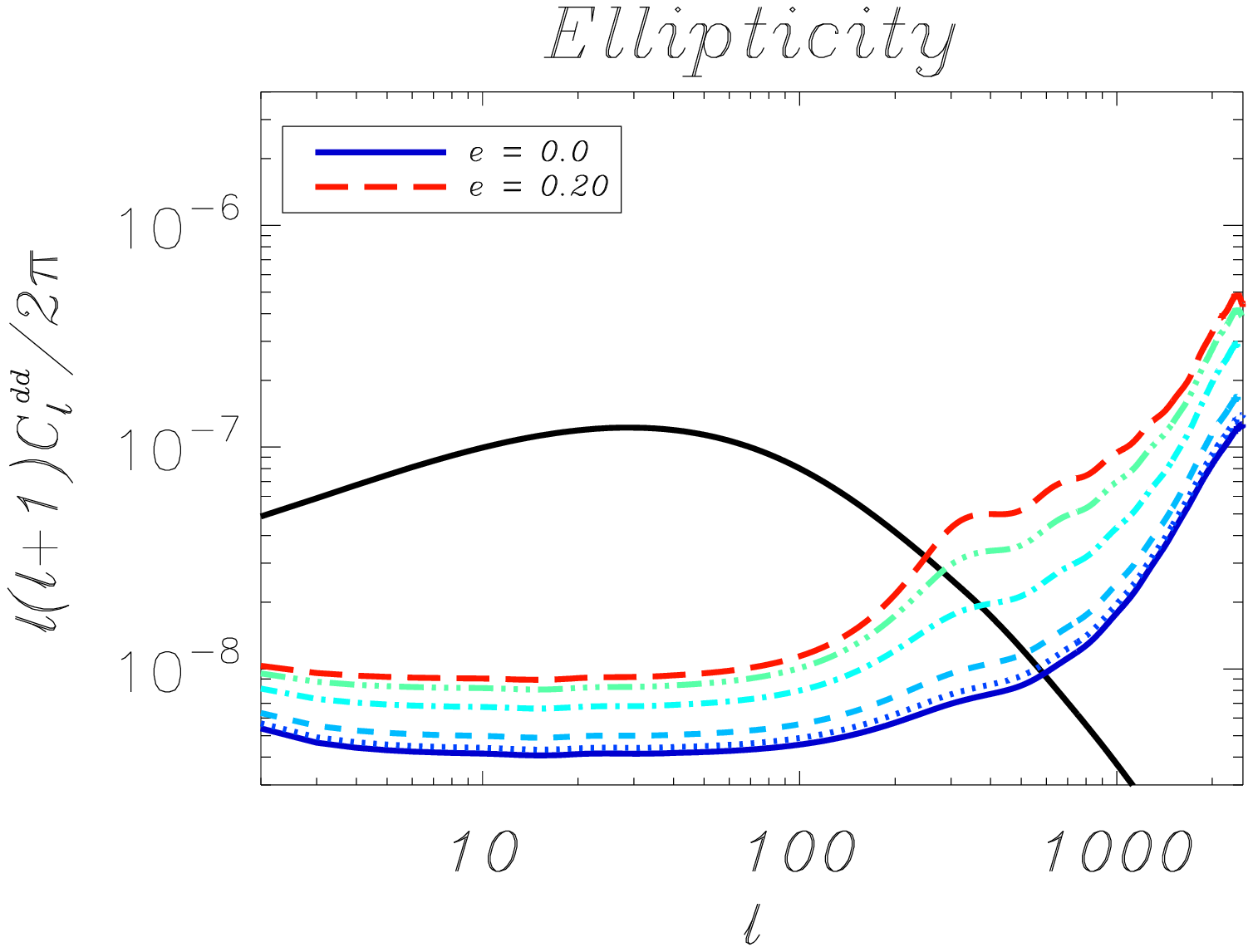} &
\includegraphics[width=\columnwidth]{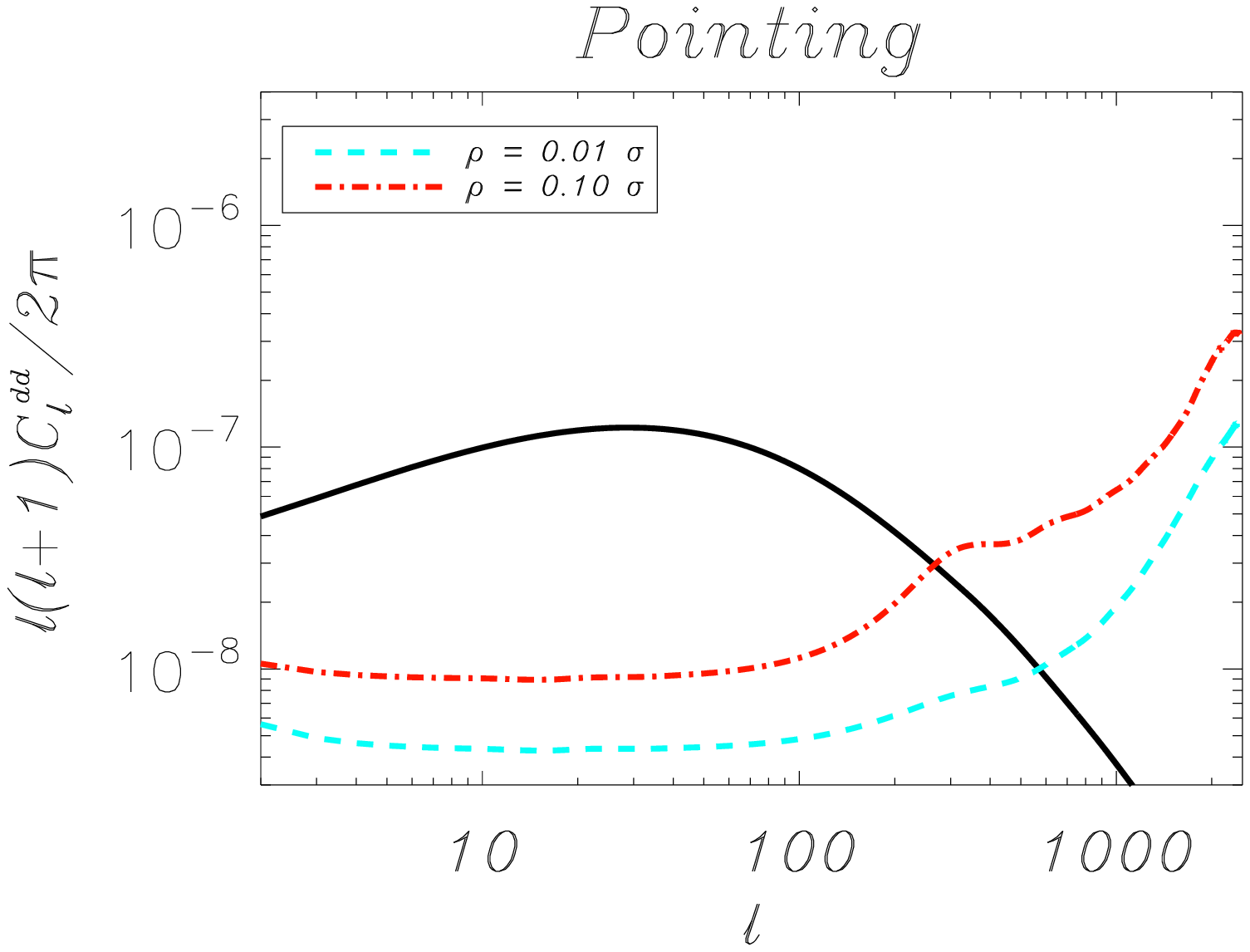}
\end{tabular}
\caption{Lensing reconstruction with CMBPOL-A: As in Figure 1.
For CMBPOL-A sensitivity and angular resolution, the  
lowest-noise estimator comes from correlations of the EB estimator. 
Therefore, lensing reconstruction is only mildly affected by beam systematics.}
\end{figure*}  

\begin{figure*}
\begin{tabular}{cc}
\includegraphics[width=\columnwidth]{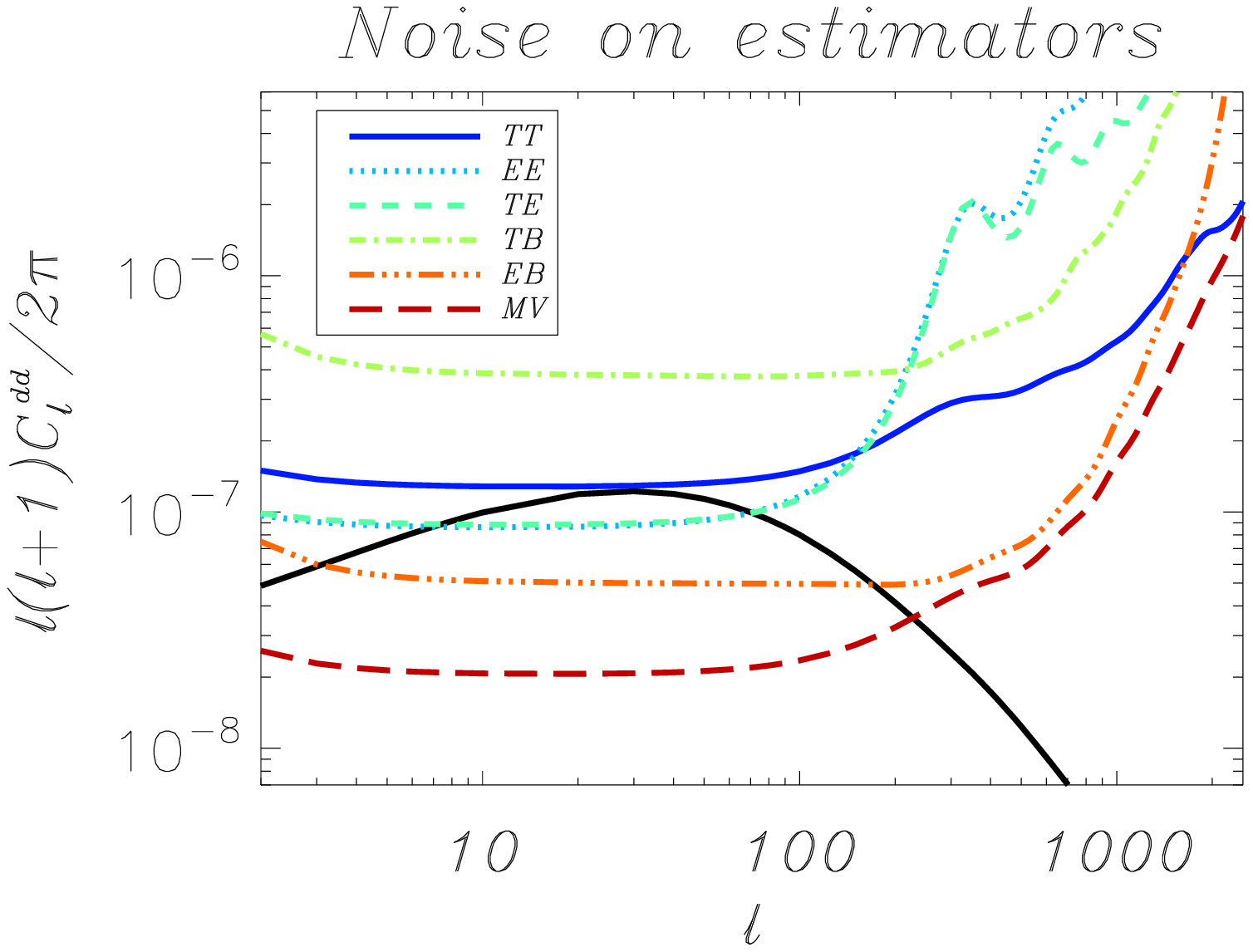} &
\includegraphics[width=\columnwidth]{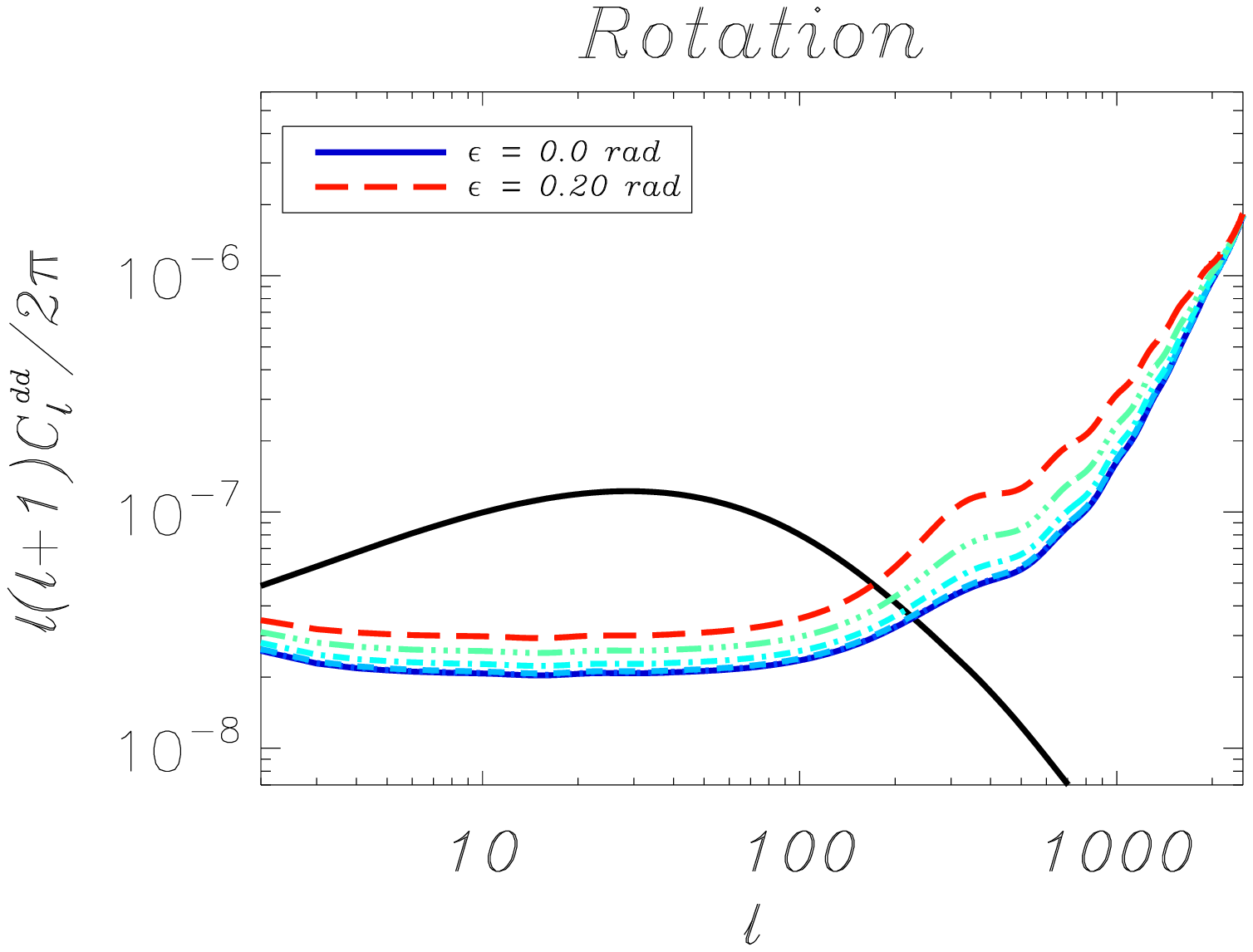} \\
\includegraphics[width=\columnwidth]{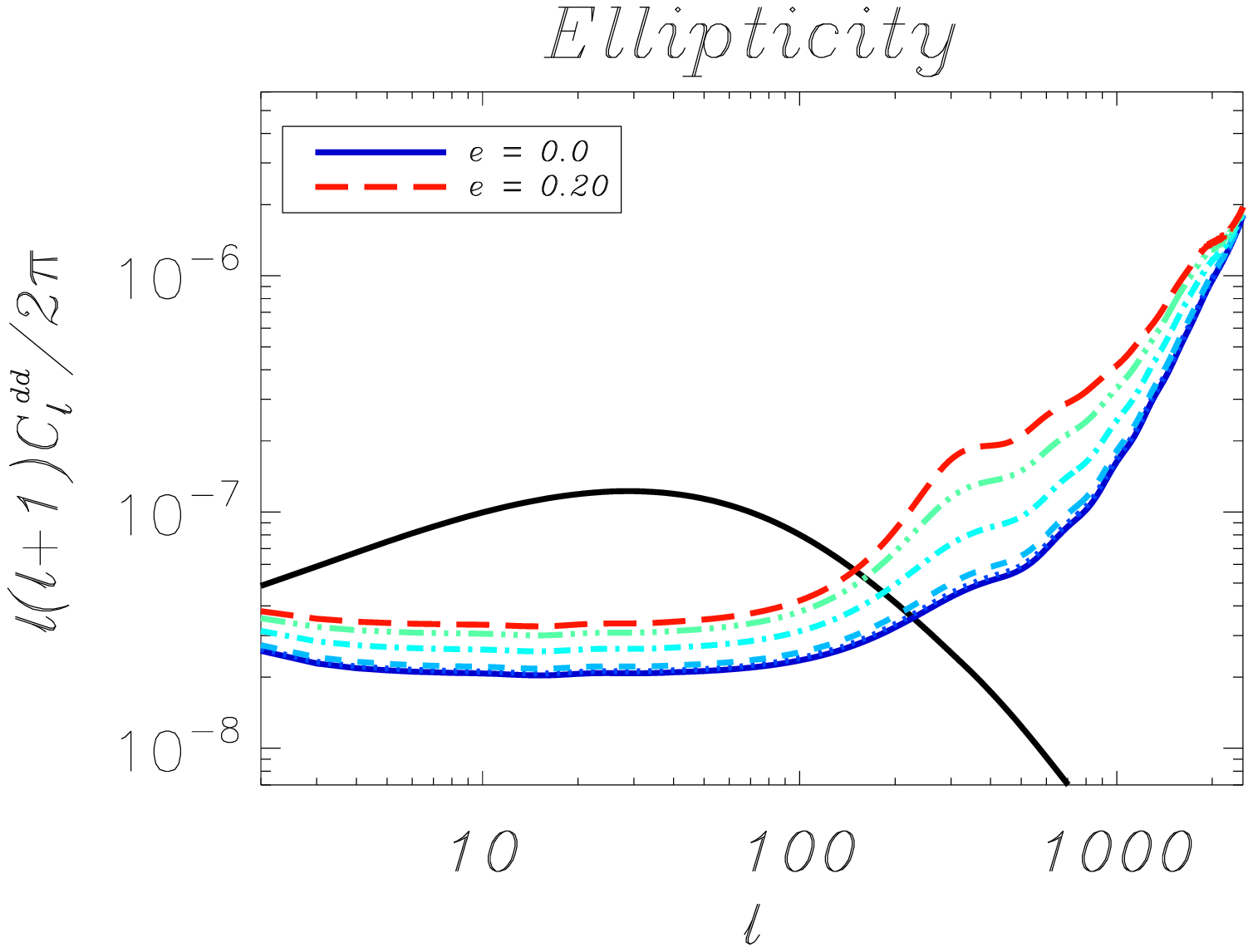} &
\includegraphics[width=\columnwidth]{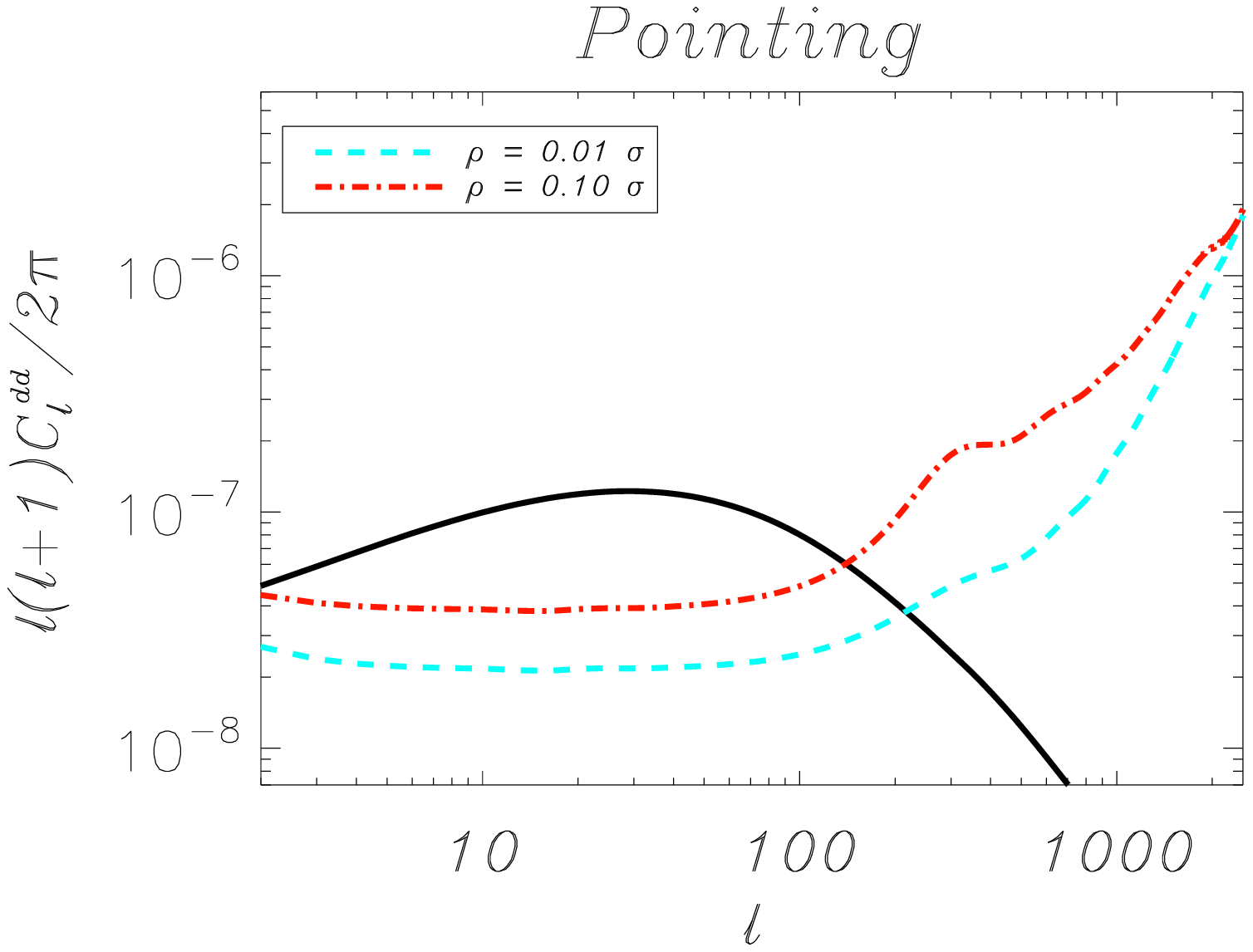}
\end{tabular}
\caption{Lensing reconstruction with QUIET+CLOVER: As in Fig.1.
For QUIET+CLOVER sensitivity and angular resolution, the  
lowest-noise estimator comes from correlations of the EB estimator. 
Therefore, lensing reconstruction is significantly affected by beam systematics.}
\end{figure*}  

\begin{figure}
\includegraphics[width=8cm]{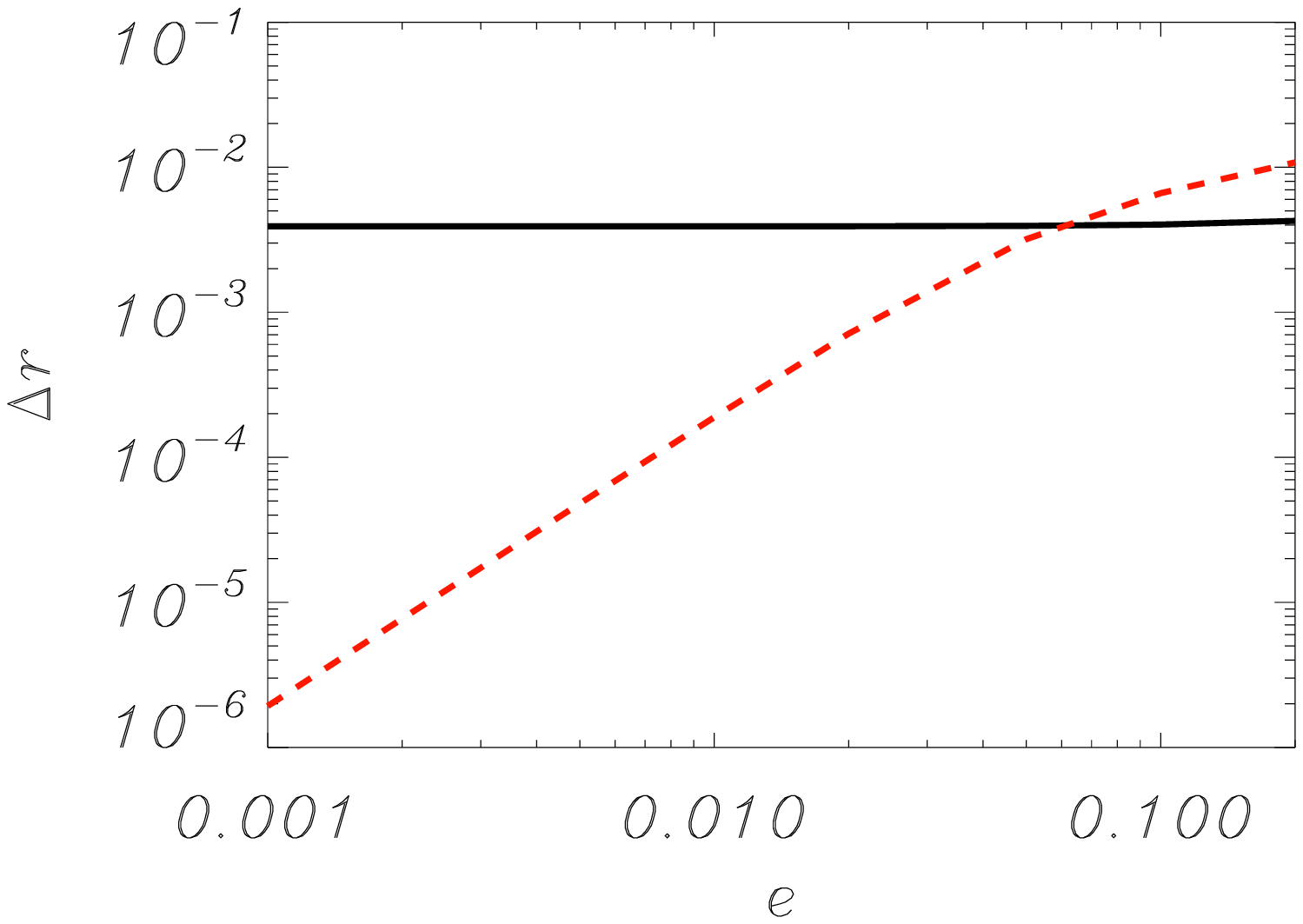}
\includegraphics[width=8cm]{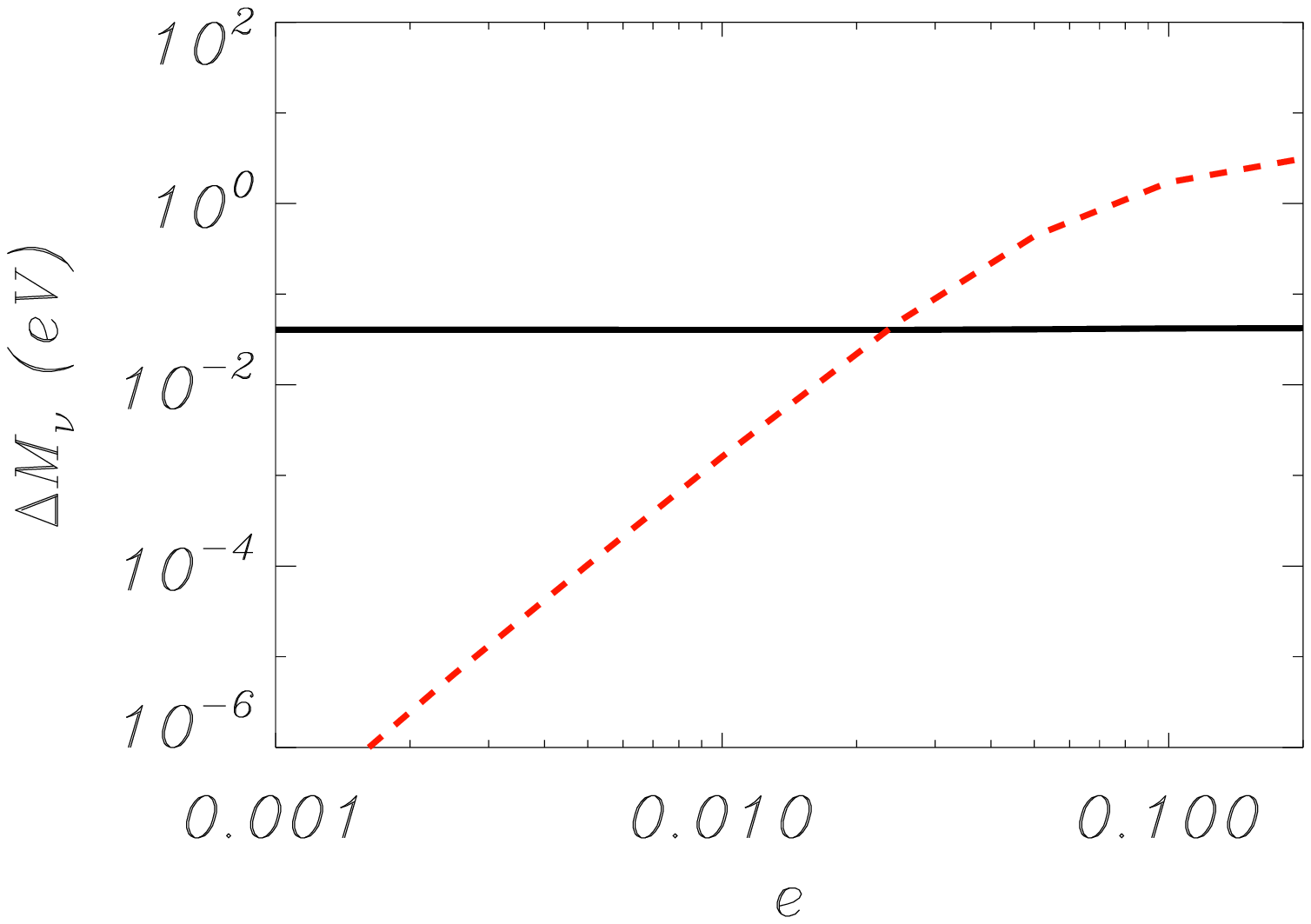}
\caption{Uncertainty in the tensor-to-scalar ratio $r$ (top) 
and total neutrino mass $M_{\nu}$ (bottom) 
due to beam ellipticity of POLARBEAR. The 
black solid curve is the statistical error (uncertainty) and the 
red dashed curve is the bias. 
As we vary $e$, the uncertainty increases by only a few percent 
(i.e. the width of the corresponding 1-D likelihood function does not 
significantly change). 
The bias, however, sharply rises with increasing ellipticity, 
i.e. the expectation value of $r$ and $M_{\nu}$ significantly 
changes. In general, we find that beam systematics 
mainly bias the inferred parameters since, for large enough 
beam systematics parameters, the spurious 
polarization signal overwhelms the cosmological signal.}
\end{figure}  

\section{Error Forecast}

Accounting for beam systematics in both Stokes parameters and lensing 
power spectra is straightforward. In addition, the instrumental noise associated with 
the main beam is accounted for, as is conventional, by adding 
an exponential noise term. Assuming gaussian white noise
\begin{eqnarray}
N_{l}=\frac{1}{\sum_{a}(N_{l}^{aa})^{-1}}
\end{eqnarray}
where $a$ runs over the experiment's frequency bands. The noise in channel {\it a} is 
(assuming a gaussian beam) 
\begin{eqnarray}
N_{l}^{aa}=(\theta_{a}\Delta_{a})^{2}e^{l(l+1)\theta_{a}^{2}/8\ln(2)},
\end{eqnarray}
where $\Delta_{a}$ is the noise per pixel in $\mu$K-arcmin,
$\theta_{a}$ is the beam width (see Table III), 
and we assume noise from different channels is uncorrelated.
The power spectrum then becomes
\begin{eqnarray}
C_{l}^{X}\rightarrow C_{l}^{X}+N_{l}^{X}
\end{eqnarray}
where $X$ is either the auto-correlations $TT$, $EE$ and $BB$ or the cross- 
correlations $TE$, $TB$ and $EB$ (the latter two power spectra vanish in the 
standard model but not in the presence of beam systematics and exotic parity-violating physics 
e.g. Carroll [23], Liu, Lee \& Ng [24], Xia et al. [25], Komatsu et al. [26] or
primordial magnetic fields e.g. Kosowsky \& Loeb [27]). For the cross-correlations, the $N_{l}^{X}$ vanish as 
there is no correlation between the instrumental noise of the 
temperature and polarization (in the absence of beam systematics).

\subsection{Fisher-Matrix-Based Calculation}

The effect of instrumental noise is simply to increase 
the error bars, which is evident from the Fisher matrix formalism below.
The 1-$\sigma$ error $\sigma(\lambda_{i})$ on the cosmological parameter $\lambda_{i}$ 
can be read-off from the appropriate diagonal element of the inverse Fisher 
matrix
\begin{eqnarray}
\sigma(\lambda_{i})=\sqrt{(F^{-1})_{ii}} \label{fisherr} 
\end{eqnarray}
where the Fisher matrix elements are defined as
\begin{eqnarray}
F_{ij}=-\Big\langle\frac{\partial^{2}L}{\partial\lambda_{i}\partial\lambda_{j}}\Big\rangle, \label{fishmat}
\end{eqnarray}
$L$ is the likelihood function, and Eq. (\ref{fishmat}) is evaluated at the best-fit point 
in parameter space.
Explicitly, the Fisher matrix elements for the CMB read
\begin{eqnarray}
F_{ij}=\frac{1}{2}\sum_{l}(2l+1)f_{{\rm sky}}
{\rm Trace}\left[{\bf C}^{-1}\frac{\partial{\bf C}}{\partial\lambda_{i}}{\bf C}^{-1}\frac{\partial{\bf C}}{\partial\lambda_{j}}\right]. \label{fishcmb}
\end{eqnarray}
The pre-factor $\frac{1}{2}(2l+1)f_{{\rm sky}}$ comes from the sample-variance of the multipole 
$l$ with an experiment covering a fraction $f_{{\rm sky}}$ of the sky.
The matrix ${\bf C}$ is
\begin{eqnarray}
{\bf C}=\left(\begin{array}{c c c c}
C'^{TT}_{l} & C_{l}^{TE} & 0 & C_{l}^{Td}\\
C_{l}^{TE} & C'^{EE}_{l} & 0 & 0\\
0 & 0 & C'^{BB}_{l} & 0\\
C_{l}^{Td} & 0 & 0 & C'^{dd}_{l}
\end{array}\right) \label{covmat}
\end{eqnarray}
where the diagonal primed elements $C'^{XX}_{l}\equiv C_{l}^{XX}+N_{l}^{XX}$ and $X\in\{T,E,B,d\}$. 
In general, $N_{l}^{EE}=N_{l}^{BB}=2N_{l}^{TT}$. 
Note that, except for $N_{l}^{dd}$, which is not an instrumental noise 
and emerges only because of the limited reconstruction of the lensing 
potential by the quadratic estimators of Hu \& Okamoto [20], 
the instrumental noise will increase ${\bf C}$, but {\it not} its derivatives 
with respect to the cosmological parameters. This will  
increase the error on the parameter estimation as seen from 
Eqs. (\ref{fisherr}), (\ref{fishcmb}), and (\ref{covmat}).
It is merely because the instrumental noise dilutes 
the information below the characteristic beamwidth scale, 
and the error increases correspondingly. However, this {\it is not} necessarily the case 
with beam systematics since they couple to the underlying cosmological 
model, and therefore {\it do} depend on cosmological parameters. This noise due to systematics, 
$N_{l}^{{\rm sys}}$, contributes to both ${\bf C}$ {\it and} $\frac{\partial{\bf C}}{\partial\lambda_{i}}$ 
and its effect on the confidence level of parameter estimation can be, in principle, 
either a degradation or an improvement. This argument ignores potential systematic errors, i.e., 
bias (systematic shift of the average of the statistical distribution which characterizes 
certain cosmological parameters) of the recovered average values of the cosmological parameters.
Indeed, as we show 
below, the main effect of beam systematics is to bias the inferred cosmological parameters, 
especially for large beam mismatch parameters, as naively expected 
(see Fig.4 for a comparison between the bias and the uncertainty induced on the tensor-to-scalar ratio and neutrino masses by beam ellipticity).
It is important to note that, although our focus is on beam systematics and their 
effect on parameter estimation, we do not include the systematics-induced $C_{l}^{TB}$ and $C_{l}^{EB}$ in the analysis 
(Eqs. 22, 23) because our main concern is how standard data analysis pipelines will be affected by beam systematics. We defer the treatment of the more general case, which includes the parity-violating terms and their effect on constraints of beyond-the-standard-model parity-violating interactions in the primordial universe, to a future work (Shimon, Miller \& Keating [28]).

Given the beam systematics the bias of a parameter can be calculated with the Fisher Matrix. 
This has been done by O'Dea, Challinor \& Johnson [6]. The bias in a parameter $\lambda_i$, if not too large, 
is given by
\begin{eqnarray}
\Delta\lambda_{i} = \langle \lambda_i^{obs} \rangle - \langle \lambda_i^{true} \rangle = \sum_j \left( \mathbf{F}^{-1} \right) _{ij} B_j
\end{eqnarray}
where the bias vector $\mathbf{B}$ can be written as

\begin{eqnarray}
	\mathbf{B} = \sum_l (\mathbf{C}_{l}^{sys})^t \mathbf{\Xi}^{-1} \frac{\partial \mathbf{C}_{l}^{cmb}}{\partial \lambda_j}
\end{eqnarray}
and $\Xi_{ij} = \textnormal{cov}(C_l^i,C_l^j)$ and $\mathbf{C}_l$ is a vector containing all six power spectra.

\subsection{Monte Carlo Simulations}

The Fisher matrix approach is known to provide reliable approximation to the uncertainty in case of 
gaussian distributions and only a lower bound for more general distributions by virtue of the Cramer-Rao 
theorem. It can yield poor estimates, however, in cases of large biases and parameter degeneracies. 
To check for such effects in our simulations we repeated the analysis with MCMC simulations which 
make use of the full likelihood function and not only its peak value and can therefore provide reliable 
estimates of parameter errors even in the presence of large biases.  
Our simulations illustrate that even when we consider the Fisher-matrix-based results as a guide for choosing the beam parameters for the MCMC simulations, the resulting bias that the Monte-Carlo simulations predict can be larger than those found with the Fisher-matrix-based calculation (actually Eq. 24 assumes that the bias is small compared to the characteristic width of the likelihood function of the parameter in question; when this is not the case this approximation is invalid) in some cases. This important point is further elucidated in the next section.
For our Monte Carlo simulations we use a modified version 
of CosmoMC [29] which includes measurements of the 
lensing potential and its cross-correlation with the temperature anisotropy when calculating the likelihood in order to run these simulations. An eleven parameter 
model is used ($\Omega_b h^2$, $\Omega_{dm} h^2$, $\Theta$, $\tau$, $\Omega_{\nu}h^{2}$, $w$, $n_s$, $n_t$, $\alpha$, $\log (10^{10} A_s)$, and $r$). 
We ran simulations for each of the five systematic effects with noise corresponding 
to POLARBEAR, CMBPOL-B and QUIET+CLOVER experiments. While running Monte Carlo simulations is much more time consuming compared 
to the Fisher matrix approach, the error forecasts for future experiments will ultimately have to account for a potentially significant biases of the inferred cosmological parameters.

\section{Results}

For both the Fisher and MCMC methods we consider the 
effect of both irreducible and reducible systematics. By `reducible' we refer to systematics which depend on the coupling of an imperfect scanning strategy to the beam mismatch parameters. 
These can, in principle, be removed or reduced during data analysis.
This includes the differential gain, differential beamwidth and first order pointing error beam systematics. By `irreducible' we refer to those systematics that depend only on the beam mismatch parameters (to leading order). For instance, the differential ellipticity and second order pointing error persist even if the scanning strategy is ideal. For reducible systematics the scanning strategy 
is a free parameter in our analysis (under the assumption it is non-ideal, yet uniform, over the map)
and we set limits on the product of the scanning strategy (encapsulated by the $f_{1}$ and $f_{2}$ parameters) and the differential gain, beamwidth and pointing, as will be described below. 

To calculate the power spectra 
we assume the concordance cosmological model throughout; the baryon, cold dark mater, 
and neutrino physical energy densities in critical density units $\Omega_{b}h^{2}=0.021$, 
$\Omega_{c}h^{2}=0.111$, $\Omega_{\nu}h^{2}=0.006$. The latter is equivalent to a total neutrino mass $M_{\nu}=\sum_{i=1}^{3}m_{\nu,i}=$0.56eV, slightly lower than the current limit set by a joint analysis of the WMAP data and a variety of other cosmological probes (0.66eV, e.g. Spergel et al. [30]). We assume degenerate neutrino masses, i.e. all neutrinos have the same mass, 0.19 eV, for the purpose of illustration, and we do not attempt to address here the question of what tolerance levels are required to determine the neutrino hierarchy. As was shown by Lesgourgues et al. [13], the prospects for determining the neutrino hierarchy from the CMB {\it alone}, even in the absence of systematics, are not very promising. 
This conclusion may change when other probes, e.g. Ly-$\alpha$ forest, are added to the analysis.
Dark energy makes up the rest of the energy required for closure density. 
The Hubble constant, dark energy equation of state and helium fraction are, 
respectively, $H_{0}=70$ ${\rm km\ sec^{-1}Mpc^{-1}}$, $w=-1$ and $Y_{He}=0.24$. 
$h$ is the Hubble constant in 100 km/sec/Mpc units.
The optical depth to reionization and its redshift are $\tau_{re}=0.073$ and $z_{re}=12$. The normalization of the primordial power spectrum was set to $A_{s}=2.4\times 10^{-9}$ and its power law index is $n_{s}=0.947$ 
(Komatsu et al. [26]).

Since the effect of beam systematics is the focus of this paper, and because these systematics are generally manifested on scales smaller than the beamwidth (except for the effects of differential gain and rotation) we concentrate on some specific cosmological parameters which will be targeted by upcoming CMB experiments. These parameters have been chosen for our analysis because we expect them to benefit from lensing extraction or simply because they are somehow associated with small angular scales and therefore are prone to systematics on sub-beam scales. We limit our analysis to the tensor-to-scalar ratio $r$, dark energy equation of state $w$, spatial curvature $\Omega_{k}$, running of the scalar index $\alpha$ and total neutrino mass $M_{\nu}$. While $r$ is mainly constrained by the primordial B-mode signal that peaks on degree scales (and is therefore not expected to be overwhelmed by the beam systematics which peak at sub-beam scales), it is still susceptible to the tail of these systematics, extending all the way to degree scales, because of its expected small amplitude (less than $0.1\mu K$). The tensor-to-scalar ratio is also affected by differential gain and rotation which are simply rescalings of temperature anisotropy and E-mode polarization power spectra, respectively, and therefore do not necessarily peak at scales beyond the primordial signal. 

The other four parameters either determine the {\it primordial} power spectrum $P(k)$ on small angular scales (e.g. $\alpha$) or affect the lensed signal (both temperature and polarization) at late times (e.g. $M_{\nu}$, $\Omega_{k}$, $w$). Ideally, the lensing signal, which peaks at $l\approx 1000$, provides a useful handle on the neutrino mass as well as other cosmological parameters which govern the evolution of the large scale structure and gravitational potentials. However, the inherent noise in the lensing reconstruction process (Hu \& Okamoto [20]) which depends, among others, on the instrument specifications (instrumental noise and beamwidth), now depends on beam systematics as well. The systematics, however, depend on the cosmological parameters through temperature leakage to polarization, and as a result there is a complicated interplay between these signals and the information they provide on cosmological parameters. As our numerical calculations show, the effect on the inferred cosmological parameters stems from both the {\it direct effect} of the systematics on the parameters (the top-left $3\times 3$ block of the covariance matrix, Eq. 23) and the {\it indirect effect} on the noise in the lensing reconstruction, $N_{l}^{dd}$, in cases where the MV estimator is dominated by the EB correlations (see section 3).

\subsection{Fisher Matrix Results}

The Fisher information-matrix gives a first order approximation to the lower bounds on errors inferred for these parameters. However, by construction, it uses only the information from the peak of the likelihood function. Markov Chain Monte Carlo simulations are known to be superior to Fisher-matrix-based analysis in cases of strong parameter degeneracies and bias but Fisher matrix results are useful for first order approximation and provide starting values for MCMC simulations. 

We follow O'Dea, Challinor \& Johnson [6] in quantifying the required tolerance on the differential gain, differential beamwidth, pointing, ellipticity and rotation. To estimate the effect of systematics and to set the systematics to a given tolerance limit one has to compare the systematics-free $1\sigma$ error in the i-th parameter (Eq.20) to the error obtained in the presence of systematics. The latter has two components; the bias and the uncertainty (which depends on the curvature of the likelihood function, i.e. to what extent does the information matrix constrain the cosmological model in question). As in O'Dea, Challinor \& Johnson [6] we define
\begin{eqnarray}
\delta &=&\frac{\Delta\lambda_{i}}{\sigma_{\lambda_{i}}}|_{\lambda_{i}^{0}}\nonumber\\
\beta &=&\frac{\Delta\sigma_{\lambda_{i}}}{\sigma_{\lambda_{i}}}|_{\lambda_{i}^{0}} 
\end{eqnarray}
where the superscript $0$ refers to values evaluated at the peak of the likelihood function, i.e. the values we assume for the underlying model, and $\Delta\lambda_{i}$ and $\Delta\sigma_{\lambda_{i}}$ are the bias (defined in Eq. 24) and the change in the statistical error for a given experiment and for the parameters $\lambda_{i}$ induced by the beam systematics, respectively. As shown in O'Dea, Challinor \& Johnson [6] these two parameters depend solely on the primordial, lensing and systematics power spectra. We require both $\delta$ and $\beta$ not to exceed 10\% of the uncertainty without systematics. As illustrated in Fig.4 for the case of tensor-to-scalar ratio and neutrino total mass, the bias exceeds the 
uncertainty at some value of the beam ellipticity. This is a general result; for given beam systematic and a cosmological parameter the bias becomes the dominant component of the error in parameter estimation for sufficiently large beam imperfection (ellipticity, gain, etc).
This sets the limit on our five systematics parameters as demonstrated in Tables IV, V, VI, VII, VIII and IX for PLANCK, POLARBEAR, SPIDER, QUIET+CLOVER, CMBPOL-A and CMBPOL-B 
(we considered two cases which we refer to as CMBPOL-A and CMBPOL-B, the former is a high sensitivity experiment with 1000 Planck-equivalent detectors, the later is motivated by Kaplinghat, Knox \& Song [12]) whose specifications are given in Table III. For the systematics power spectra we used the expressions in Table II assuming only temperature leakage (i.e. polarization-free underlying sky) except for the effect of differential rotation where we consider mixing between the E- and B-mode. $C^{TE}$ cannot leak to the B-mode power spectrum since it assumes negative values for certain multipole numbers and $C^{E}$-contribution is higher order correction to the B-mode systematics and will add only few percent at most to the induced systematics. Due to the scaling of the systematics with the beam width, this potentially negligible $C^{E}$ contribution, which contaminate the B-mode polarization at second order, will result in only a few percent change to the tolerance levels for the beam parameters we consider. 
More specifically, Tables IV, V, VI, VII, VIII and IX contain the maximum differential gain, beamwidth, pointing, ellipticity and rotation, in units of $\sqrt{\frac{f_{1}}{2\pi}}$ of a percent, $\sqrt{\frac{f_{2}}{2\pi}}$ of an arcsecond, a percent and a degree, respectively. As explained in Eq.(12) the factor $2\pi$ in the denominators is the result of our assumption that the scanning strategy is spatially uniform.
All results are robust against changing the step size used.
Under the assumption of uniform scanning strategy and assuming `worst case scenario' that the quadrupole moment of the scanning strategy is maximal ($\sqrt{\frac{f_{1}}{2\pi}}=1$) in case of differential gain and beamwidth and that the monopole and octupole moments are maximal ($\sqrt{\frac{f_{2}}{2\pi}}=1$) these allowed values correspond directly to $g$ and $\mu$ in (percent units) and $\rho$ (in arcsec units), respectively.
It is also apparent from the tables that better sensitivity and higher resolution experiments generally require better control of beam parameters. This is expected since our criterion is the {\it fixed} 10\% threshold in $\beta$ and $\delta$, i.e. the systematics are not allowed to exceed the 10\% (or any other reasonably chosen threshold) level in the parameter uncertainty ($\sigma_{\lambda_{i}}$) units. Higher sensitivity experiments will have smaller 
$\sigma_{\lambda_{i}}$ in general and therefore the allowed $g$, $\mu$, $e$, $\rho$ and $\varepsilon$ will be smaller. 
This does not imply necessarily that controlling beam systematics of higher sensitivity experiments will be more challenging since the uncertainty of beam parameters is a direct result of the S/N level with which the beam is calibrated against a point-source. Reducing the detector noise (akin to higher-sensitivity experiment) allows smaller uncertainty in beam parameters. Also, as mentioned above the 10\% threshold adapted here is arbitrarily chosen and as long as we keep the systematic bias on the cosmological parameters $\lambda_{i}$ smaller than $\sigma_{\lambda_{i}}$, e.g. even in case $\delta$ is as high as 0.2, the beam systematics will not significantly degrade the science. Therefore, even very sensitive high-resolution experiments are expected to yield good systematics control.

\subsection{MCMC Results}

This work is the first to employ Monte-Carlo simulations to assess the effect 
of beam systematics on parameter estimation. We first considered beam parameters 
which bias the tensor-to-scalar ratio by 10\%, 50\% and 100\% of 
the error (i.e. $\delta =0.1$, $0.5$ and $1.0$, see Eq. 26) found with the Fisher matrix formalism. The 
limiting value of $\delta=0.1$ was chosen for the Fisher matrix formalism so that it was small enough such that we should not be able to see a bias, 
which is what we want for a limit. For the MCMC simulations, we want to be able to distinctly see a bias and with $\delta=1$ we expected to observe it.
The other two values $\delta=0.1$ and $0.5$ are also reported mainly for the purpose of comparison between the naive expectations (based on Fisher matrix analysis) with the MCMC results.
For these simulations we focus only on the most sensitive cosmological parameters: $r$, $w$ and $M_{\nu}$. We ran the MCMC analysis on the experiments 
POLARBEAR, CMBPOL-B, and QUIET+CLOVER. 

Our results for POLARBEAR are reported in Table X, for CMBPOL-B in Table XI, and for QUIET+CLOVER in Table XII. 
The most sensitive parameter to the 
beam systematics considered here turns out to be $r$, the tensor-to-scalar ratio as it is not constrained by the addition of the lensing power spectra.
We found the bias on $r$ can be as high as $\sim 100\%$. $w$ and $M_{\nu}$, which are mostly constrained by the larger 
(compared to the primordial signal from inflation) lensing signal, are changed by no more that $\sim 10\%$, itself a non-negligible bias. Even in the absence of systematics POLARBEAR and QUIET+CLOVER exhibit a small bias (approximately $\frac{1}{2}\sigma$) in $w$ (Tables X and XII, respectively) towards values smaller than -1 but this situation significantly improves with CMBPOL-B. The reason is that, as is evident from our simulation, the 1-D distribution for $w$, while peaked at $-1$, is skewed towards more negative values. As the experiment sensitivity improves, such as in CMBPOL-B, this small bias becomes insignificant.
Most importantly, we also found that the levels of bias (in $r$) caused by 
differential beamwidth and ellipticity exceed the bias found with the naive inclusion of the power spectrum bias in the Fisher-matrix formalism (Table XIII). 
This illustrates that the simplistic approach to bias within the Fisher-matrix formalism 
{\it underestimates} the induced bias on the cosmological parameters. However, the Fisher matrix can be used, as done here, to determine the starting values for MCMC simulations. 
These two systematics, the differential beamwidth and ellipticity, are {\it second} order gradients of the underlying temperature anisotropy as opposed 
to the first order gradient in case of, e.g. first order pointing error effect. This implies that for given $\mu$ and $e$ this effect steeply 
increases towards smaller scales. The Fisher-matrix bias calculation is based, however, on the assumption that the bias is relatively small, an assumption 
which certainly breaks down when high resolution experiments are considered (i.e. with SPIDER's comparatively low angular resolution, for example, we 
expect the tension between the Fisher-matrix-based and MCMC estimations of the bias to be smaller). 

\section{Conclusions}

The purpose of this work was to illustrate the effect of beam systematics on parameter extraction from CMB observations. Beam systematics are expected to be significant especially for detecting the B-mode polarization.
Ongoing and future experiments must meet very challenging requirements at the experiment design and data analysis phases to assure polarimetric fidelity. Ultimately, a major target of these experiments is the most accurate estimation of cosmological parameters, and for this end it is mandatory to assess, among other issues, the propagation of beam systematics to parameter estimation. The tolerance levels chosen in this work are somewhat arbitrary and may be changed at will, according to the goals of individual experiments, and the numerical values we quote in the tables should be viewed in this perspective.

The only similar work so far to set tolerance levels on beam mismatch in the context 
of parameter estimation is O'Dea, Challinor \& Johnson [6] which influenced our present work.
However, we expand on this work in several ways. While O'Dea, Challinor \& Johnson [6] 
considered only the effect of systematics on the tensor-to-scalar ratio $r$, 
we consider a family of parameters associated with the B-mode sector: 
$r$, $M_{\nu}$, $\alpha$, $w$ and $\Omega_{k}$.
We set all other cosmological parameters to be consistent with the WMAP values.
In order to exhaust the potential of the CMB to constrain these parameters 
we carried out lensing extraction. In addition, we repeated the analysis 
for POLARBEAR, CMBPOL-B and QUIET+CLOVER with Monte Carlo simulations and 
found that the Fisher-Matrix approximation is, in general, inadequate for appraising the biases. 
We also found that high resolution experiments, such as 
POLARBEAR are very sensitive to bias from second order gradient effects (i.e. differential 
ellipticity and differential beamwidth) which is underestimated by the Fisher-matrix-based calculation, but fully accounted for with MCMC simulations. Also, unlike O'Dea, Challinor \& Johnson [6] our results 
are presented independently of the scanning strategy details. The only assumption we 
made was that the scanning strategy is spatially uniform, a condition which can be 
achieved with or without a HWP which samples the polarization angles in a way which 
is uniform; both spatially, and in terms of polarization angle.
In case that this approximation fails the more general formalism 
(Shimon et al. [7]) should be used with the added complexity introduced to lensing reconstruction 
by the scanning-induced non-gaussianity of the systematic B-mode.

We find that parameter bias is the dominant factor and its level actually sets the 
upper bounds on the beam parameters appearing in Tables IV through IX. Our results show that  
the most severe constraints are set on the most sensitive experiments for a given tolerance on 
$\delta$ and $\beta$ since these quantities are experiment-dependent (Eq. 26) and since, 
in general, an experiment with higher resolution and better sensitivity will result in 
smaller errors $\sigma_{\lambda_{i}}$. We expect that the constraints on the systematics should be 
more demanding so as to realize the potential of experiments. 
As mentioned above, the most stringent constraints are obtained from the requirement on the bias rather than from increased parameter uncertainty. Again, for the same reason, as shown for specific examples in Fig. 4; the bias {\it always} exceeds the uncertainty for large enough systematics and this always takes place before the $10\%$ thresholds in Eqs.(26) are attained. The reason is that for large enough systematics the induced spurious polarization becomes comparable to, or exceeds, the underlying polarization signals, therefore biasing the deduced value. It is easy to visualize configurations in which the bias increases without bound while the `curvature' of the likelihood function (i.e. the statistical error) with respect to specific cosmological parameters does not change. 
It is also clear from the tables that, in general, the tensor-to-scalar ratio is the most sensitive parameter, and the second most sensitive is $\alpha$, the running of the scalar index (although there are some exceptions). If the tensor-to-scalar ratio is larger than the case we studied ($r=0.01$), this conclusion may change since $r$ is mainly affected by the overwhelming B-mode systematics on degree scales.
$\alpha$ is predicted to vanish by the simplest models of inflation and was added to parameter space to better fit the WMAP and other cosmological data. 
As is well-known, information from Ly-$\alpha$ systems and other LSS probes 
can, in principle, better constrain $\alpha$ if their associated systematics can be controlled to a sufficiently accurate level. For these small scales the CMB is not the ideal tool to extract information and the error that beam systematics induce on $\alpha$ are not significant. 

The upper limits we obtained in this work on the allowed range of beam mismatch parameters for given experiments and given arbitrarily-set tolerance levels on the parameter bias and uncertainty, constitute very conservative limits. It can certainly be the case that some of the systematics studied here may be fully or partially removed. This includes, in particular, the first order pointing error which couples to the dipole moment of non-ideal scanning strategies (see Shimon et al. [7]). By removing this dipole during data analysis the effect due to the systematic first order pointing error (dipole) drops dramatically. We made no attempt to remove or minimize these effects in this work. Our results highlight the need for scan mitigation techniques because the coupling of several beam systematics to non-ideal scanning strategies results in systematic errors. This potential solution reduces systematics, which ultimately propagate to parameter estimation, and affect mainly the parameters considered in this work. A brute-force strategy to idealize the data could be to remove data points that contribute to higher-than-the-monopole moments in the scanning strategy.
This would effectively make the scanning strategy `ideal' and alleviate the effect of the {\it a priori} most pernicious beam systematics. 
This procedure `costs' only a minor increase in the instrumental noise (due to throwing out a fraction of the data) 
but will greatly reduce the most pernicious reducible beam systematic, i.e. the first order pointing error (`dipole' effect). 
The lesson is clear: the rich treasures of cosmological parameters deducible from B-mode data  require a combination of high polarimetric fidelity and judicious data mining. Both are eminently feasible upcoming 
CMB polarization experiments. 

\section*{Acknowledgments}

We acknowledge the use of the publically available code by Lesgourgues, Perotto, Pastor \& Piat for the calculation of the noise in lensing reconstruction. We also used CAMB and CosmoMC for calculations of the Fisher matrices and the Monte Carlo simulations. 
We acknowledge using resources of the San Diego Supercomputing Center (SDSC).
We thank Oliver Zahn for critically reading this paper and for his 
very useful constructive suggestions.
Useful correspondence with Anthony Challinor is gratefully acknowledged.
BK gratefully acknowledges support from NSF PECASE Award AST-0548262.

\begin{table*}[c]
\begin{tabular}[c]{|c|c|c|c|}
\hline
~{\rm depends on} ~&~effect~&~parameter~&~definition~\\
~{\rm beam substructure}~&~~&~~&~~\\
\hline
~{\rm No}~&~gain~& $g$ & $g_{1}-g_{2}$ \\
\hline
~{\rm Yes}~&~monopole~& $\mu$ & $\frac{\sigma_{1}-\sigma_{2}}{\sigma_{1}+\sigma_{2}}$ \\
\hline
~{\rm Yes}~&~dipole~& $\rho$ & ${\bf \rho}_{1}-{\bf \rho}_{2}$ \\
\hline
~{\rm Yes}~&~quadrupole~& $e$ & $\frac{\sigma_{x}-\sigma_{y}}{\sigma_{x}+\sigma_{y}}$ \\
\hline
~{\rm No}~&~rotation~& $\varepsilon $ & $\frac{1}{2}(\varepsilon_{1}+\varepsilon_{2})$ \\
\hline
\end{tabular}
\caption{Definitions of the parameters associated with the systematic effects.
Subscripts 1 and 2 refer to the first and second polarized beams of the dual beam polarization assumed in this work.} 
\end{table*}

\begin{table*}[c]
\begin{tabular}{|c|c|c|c|c|c|}
\hline
~effect~&~parameter~&~$\Delta C_{l}^{TE}$~&~$\Delta C_{l}^{E}$~&~$\Delta C_{l}^{B}$~\\
\hline
~gain~& $g$ & 0 & $g^{2}f_{1}\star C_{l}^{T}$ & $g^{2}f_{1}\star C_{l}^{T}$\\
\hline
~monopole~& $\mu$ & 0 & $4\mu^{2}(l\sigma)^{4}C_{l}^{T}\star f_{1}$ & $4\mu^{2}(l\sigma)^{4} C_{l}^{T}\star f_{1}$\\
\hline
~pointing~& $\rho$ & $-c_{\theta}J_{1}^{2}(l\rho)C_{l}^{T}\star f_{3}$ & 
$J_{1}^{2}(l\rho)C_{l}^{T}\star f_{2}$ & $J_{1}^{2}(l\rho)C_{l}^{T}\star f_{2}$\\
\hline
~quadrupole~& e & $-I_{0}(z)I_{1}(z)c_{\psi}C_{l}^{T}$ & 
$I_{1}^{2}(z)c_{\psi}^{2}C_{l}^{T}$ & $I_{1}^{2}(z)s_{\psi}^{2}C_{l}^{T}$ \\
\hline
~rotation~& $\varepsilon$ & $0$ & $4\varepsilon^{2}C_{l}^{B}$ & $4\varepsilon^{2}C_{l}^{E}$\\
\hline
\end{tabular}
\caption{The scaling laws for the systematic effects to 
the power spectra $C_{l}^{T}$, $C_{l}^{TE}$, $C_{l}^{E}$ and $C_{l}^{B}$ 
assuming the underlying sky is not polarized (except 
for the {\it rotation} signal where we assume the E, and B-mode signals are present) 
and a general, not necessarily ideal or uniform, scanning strategy. The next order contribution 
($~10\%$ of the `pure' temperature leakage shown in the table) is contributed by $C_{l}^{TE}$. 
It can be easily calculated based on the general expressions in Shimon et al. [7] where 
the definitions of $z$, $\rho$, $\varepsilon$, etc., are also found. 
For the pointing error we found that the `irreducible' contribution to B-mode contamination, arising from a second order effect, is extremely small and therefore only the first order terms (which vanish in ideal scanning strategy) are shown. The functions $f_{1}$ and $f_{2}$ are experiment-specific and encapsulate the information about the scanning strategy which couples to the beam mismatch parameters to generate spurious polarization. In general, the functions $f_{1}$ and $f_{2}$ are spatially-anisotropic but for simplicity, and to obtain a first-order approximation, we consider them  constants (see sec. 2.2) in general. In the case of ideal scanning strategy they identically vanish. The exact expressions are given in Shimon et al. [7].} 
\end{table*}
 
\begin{table*}[c]
\begin{tabular}{|c|c|c|c|c|c|}
\hline
Experiment & $f_{\rm sky}$ & $\nu$ & $\theta_b$ & $\Delta_T$ & $\Delta_E$\\
\hline
{\sc PLANCK} & 0.65
    &  30 & 33'  &  2.0 &  2.8\\
&   &  44 & 24'  &  2.7 &  3.9\\
&   &  70 & 14'  &  4.7 & 6.7\\
&   & 100 & 9.5' &  2.5 & 4.0\\
&   & 143 & 7.1' &  2.2 & 4.2\\
&   & 217 & 5.0' &  4.8 & 9.8\\
&   & 353 & 5.0' & 14.7 & 29.8\\
&   & 545 & 5.0' &  147  & $\infty$\\
&   & 857 & 5.0' &  6700 & $\infty$\\
\hline
POLARBEAR & 0.03
  & 90 & 6.7' & 1.13 & 1.6\\
& & 150 & 4.0' & 1.70 & 2.4\\
& & 220 & 2.7' & 8.0 & 11.3\\
\hline
SPIDER & 0.6
  & 96 & 58' & 0.46 & 0.65\\
& & 145 & 40' & 0.50 & 0.71\\
& & 225 & 26' & 2.22 & 3.14\\
& & 275 & 21' & 5.71 & 8.08\\
\hline
QUIET+CLOVER & 0.015
  & 150 & 10' & 0.34 & 0.48\\
\hline
CMBPOL-A     & 0.65 &  150 & 5' & 0.22 & 0.32 \\
\hline
CMBPOL-B     & 0.65 &  150 & 3' & 1.0 & 1.4 \\
\hline

\end{tabular}
\caption{
Instrumental characteristics of the CMB polarimeters considered in this work:
$f_{\rm sky}$ is the observed fraction of the sky,
$\nu$ is the center frequency of the channels in GHz,
$\theta_b$ is the FWHM (Full-Width at Half-Maximum) in arc-minutes,
$\Delta_{T}$ is the temperature
sensitivity per pixel in $\mu$K and $\Delta_E=\Delta_B$ is the
polarization sensitivity. For all experiments, we
assumed one year of observations 
(PLANCK [32], POLARBEAR [33], SPIDER [18], QUIET+CLOVER [6]).
CMBPOL A \& B represent toy experiments for illustration with CMBPOL-A having 
1000 PLANCK detectors and PLANCK resolution and CMBPOL-B has higher resolution 
but 1$\mu K$ noise level (Kaplinghat, Knox \& Song [12]).}
\end{table*}

\begin{table*}[c]
\begin{tabular}{|c|c|c|c|c|c|c|}
\hline
~{\rm Parameter}~&~{\rm Nominal value}~&~$\frac{g}{1\%}\sqrt{\frac{f_{1}}{2\pi}}$~&~$\frac{\mu}{1\%}\sqrt{\frac{f_{1}}{2\pi}}$~&~$(\frac{\rho}{1"})\sqrt{\frac{f_{2}}{2\pi}}$~&~$\frac{e}{1\%}$~&~$\varepsilon\ [{\rm deg}]$~\\
\hline
~$r$~&~0.01~&~0.02~&~0.42~&~1.5~&~0.8~&~0.72~\\
\hline
~w~&~-1~&~0.33~&~0.38~&~2.5~&~2.4~&~2.86~\\
\hline
~$\Omega_{k}$~&~0~&~0.37~&~0.44~&~3.0~&~2.6~&~3.72~\\
\hline
~$\alpha$~&~0~&~0.67~&~0.33~&~2.2~&~2.1~&~2.23~\\
\hline
~$M_{\nu}[{\rm eV}]$~&~0.56~&~0.32~&~0.38~&~2.4~&~2.4~&~2.58~\\
\hline
\end{tabular}
\caption{Systematics tolerance for PLANCK: shown are the nominal cosmological parameters we used 
along with the tolerance levels (as defined by the criterion that both $\delta$ and $\beta$, Eq. 26, 
should not exceed the $10\%$ threshold)
set on combinations of the quadrupole of the scanning strategy 
($f_{1}$, under the assumption of uniform scanning strategy) and the dimensionless differential gain $g$ 
and differential beamwidth $\mu$. Also is shown the constraint on pointing error weighted by 
the dipole of the scanning strategy ($f_{2}$) in arcsec units. The tolerance level on ellipticity 
$e$ is dimensionless (we assumed the worst-case-scenario that $\psi=45^{\circ}$) and the allowed rotation $\varepsilon$ is given in angular degrees. 
Except for the differential beamwidth effect, the most severe constraints are obtained from the requirement 
that $r$ is not biased. $g$ and $\mu$, the parameters representing the differential gain and differential beamwidth, are defined in Table I.} 
\end{table*}

\begin{table*}[c]
\begin{tabular}{|c|c|c|c|c|c|c|}
\hline
~{\rm Parameter}~&~{\rm Nominal value}~&~$\frac{g}{1\%}\sqrt{\frac{f_{1}}{2\pi}}$~&~$\frac{\mu}{1\%}\sqrt{\frac{f_{1}}{2\pi}}$~&~$(\frac{\rho}{1"})\sqrt{\frac{f_{2}}{2\pi}}$~&~$\frac{e}{1\%}$~&~$\varepsilon\ [{\rm deg}]$~\\
\hline
~$r$~&~0.01~&~0.01~&~0.74~&~0.5~&~1.4~&~0.25~\\
\hline
~w~&~-1~&~0.16~&~0.38~&~1.7~&~1.8~&~1.26~\\
\hline
~$\Omega_{k}$~&~0~&~0.18~&~0.39~&~1.8~&~1.8~&~2.01~\\
\hline
~$\alpha$~&~0~&~0.17~&~0.30~&~1.2~&~1.3~&~0.77~\\
\hline
~$M_{\nu}[{\rm eV}]$~&~0.56~&~0.15~&~0.42~&~1.9~&~1.8~&~1.06~\\
\hline
\end{tabular}
\caption{Systematics tolerance for POLARBEAR: As in Table IV.} 
\end{table*}

\begin{table*}[c]
\begin{tabular}{|c|c|c|c|c|c|c|}
\hline
~{\rm Parameter}~&~{\rm Nominal value}~&~$\frac{g}{1\%}\sqrt{\frac{f_{1}}{2\pi}}$~&~
$\frac{\mu}{1\%}\sqrt{\frac{f_{1}}{2\pi}}$~&~$(\frac{\rho}{1"})\sqrt{\frac{f_{2}}{2\pi}}$~&~$\frac{e}{1\%}$~&~$\varepsilon\ [{\rm deg}]$~\\
\hline
~$r$~&~0.01~&~0.03~&~0.10~&~2.2~&~0.19~&~0.97~\\
\hline
~w~&~-1~&~0.13~&~0.31~&~9.2~&~0.47~&~2.86~\\
\hline
~$\Omega_{k}$~&~0~&~0.10~&~0.57~&~9.9~&~1.75~&~3.43~\\
\hline
~$\alpha$~&~0~&~0.19~&~0.12~&~5.9~&~0.55~&~6.88~\\
\hline
~$M_{\nu}[{\rm eV}]$~&~0.56~&~0.10~&~0.26~&~10.9~&~0.38~&~3.72~\\
\hline
\end{tabular}
\caption{Systematics tolerance for SPIDER: As in Table IV.} 
\end{table*}

\begin{table*}[c]
\begin{tabular}{|c|c|c|c|c|c|c|}
\hline
~{\rm Parameter}~&~{\rm Nominal value}~&~$\frac{g}{1\%}\sqrt{\frac{f_{1}}{2\pi}}$~&~
$\frac{\mu}{1\%}\sqrt{\frac{f_{1}}{2\pi}}$~&~$(\frac{\rho}{1"})\sqrt{\frac{f_{2}}{2\pi}}$~&
~$\frac{e}{1\%}$~&~$\varepsilon\ [{\rm deg}]$~\\
\hline
~$r$~&~0.01~&~0.009~&~0.20~&~0.4~&~0.4~&~0.2~\\
\hline
~w~&~-1~&~0.114~&~0.17~&~3.2~&~0.6~&~0.9~\\
\hline
~$\Omega_{k}$~&~0~&~0.122~&~0.18~&~3.5~&~0.7~&~1.0~\\
\hline
~$\alpha$~&~0~&~0.148~&~0.13~&~1.4~&~0.4~&~0.6~\\
\hline
~$M_{\nu}[{\rm eV}]$~&~0.56~&~0.109~&~0.18~&~3.1~&~0.7~&~0.8~\\
\hline
\end{tabular}
\caption{Systematics tolerance for QUIET+CLOVER: As in Table IV.} 
\end{table*}

\begin{table*}[c]
\begin{tabular}{|c|c|c|c|c|c|c|}
\hline
~{\rm Parameter}~&~{\rm Nominal value}~&~$\frac{g}{1\%}\sqrt{\frac{f_{1}}{2\pi}}$~&~
$\frac{\mu}{1\%}\sqrt{\frac{f_{1}}{2\pi}}$~&~$(\frac{\rho}{1"})\sqrt{\frac{f_{2}}{2\pi}}$~&~$\frac{e}{1\%}$~&~$\varepsilon\ [{\rm deg}]$~\\
\hline
~$r$~&~0.01~&~0.0016~&~0.05~&~0.04~&~0.10~&~0.023~\\
\hline
~w~&~-1~&~0.0259~&~0.19~&~0.4~&~0.28~&~0.773~\\
\hline
~$\Omega_{k}$~&~0~&~0.0270~&~0.21~&~0.4~&~0.28~&~0.372~\\
\hline
~$\alpha$~&~0~&~0.0266~&~0.08~&~0.3~&~0.21~&~0.123~\\
\hline
~$M_{\nu}[{\rm eV}]$~&~0.56~&~0.0251~&~0.18~&~0.4~&~0.28~&~0.401~\\
\hline
\end{tabular}
\caption{Systematics tolerance for CMBPOL-A: As in Table IV.} 
\end{table*}

\begin{table*}[c]
\begin{tabular}{|c|c|c|c|c|c|c|}
\hline
~{\rm Parameter}~&~{\rm Nominal value}~&~$\frac{g}{1\%}\sqrt{\frac{f_{1}}{2\pi}}$~&~
$\frac{\mu}{1\%}\sqrt{\frac{f_{1}}{2\pi}}$~&~$(\frac{\rho}{1"})\sqrt{\frac{f_{2}}{2\pi}}$~&~$\frac{e}{1\%}$~&~$\varepsilon\ [{\rm deg}]$~\\
\hline
~$r$~&~0.01~&~0.0031~&~0.57~&~0.2~&~1.1~&~0.066~\\
\hline
~w~&~-1~&~0.0728~&~0.40~&~0.9~&~1.7~&~0.716~\\
\hline
~$\Omega_{k}$~&~0~&~0.0762~&~0.39~&~0.8~&~1.8~&~0.888~\\
\hline
~$\alpha$~&~0~&~0.0600~&~0.30~&~0.4~&~1.3~&~0.315~\\
\hline
~$M_{\nu}[{\rm eV}]$~&~0.56~&~0.0700~&~0.45~&~1.4~&~1.7~&~0.544~\\
\hline
\end{tabular}
\caption{Systematics tolerance for CMBPOL-B: As in Table IV.} 
\end{table*}

\begin{table*}[c]
\begin{tabular}{|c|c|c|c|c|c|c|c|}
\hline
~Parameter~&~{\rm no sys.}~&~$\delta^{Fish}$~&~{\rm diff. gain}~&~{\rm diff. pointing}~&~{\rm diff. beamwidth}~&
~{\rm diff. ellipticity}~&~{\rm diff. rotation}~\\
\hline
~$r$~&~$0.0103\pm 0.0036$~&~0.1~&~$0.0108\pm 0.0037$~&~$0.0108\pm 0.0038$~&~$0.0109\pm 0.0036$~
&~$0.0107\pm 0.0038$~&~$0.0109\pm 0.0037$\\
~$$~&~$$~&~0.5~&~$0.0123\pm 0.0038$~&~$0.0124\pm 0.0040$~&~$0.0133\pm 0.0042$~
&~$0.0126\pm 0.0041$~&~$0.0122\pm 0.0049$\\
~$$~&~$$~&~1.0~&~$0.0152\pm 0.0042$~&~$0.0152\pm 0.0047$~&~$0.0205\pm 0.0058$~
&~$0.0192\pm 0.0056$~&~$0.0147\pm 0.0043$\\
\hline
~$w$~&~$-1.170\pm 0.328$~&~0.1~&~$-1.170\pm 0.328$~&~$-1.168\pm 0.325$~&~$-1.217\pm 0.328$~&
~$-1.154\pm 0.334$~&~$-1.180\pm 0.335$\\
~$$~&~$$~&~0.5~&-$1.163\pm 0.307$~&~$-1.186\pm 0.331$~&~$-1.391\pm 0.269$~&~$-1.174\pm 0.329$~&~$-1.172\pm 0.327$\\
~$$~&~$$~&~1.0~&-$1.171\pm 0.321$~&~$-1.194\pm 0.334$~&~$-1.575\pm 0.217$~&~$-1.153\pm 0.333$~&~$-1.145\pm 0.320$\\
\hline
~$M_{\nu}[{\rm eV}]$~&~$0.537\pm 0.071$~&~0.1~&~$0.539\pm 0.068$~&~$0.538\pm 0.070$~&~$0.558\pm 0.068$~&
~$0.531\pm 0.069$~&~$0.540\pm 0.070$\\
~$~~$~&~$~~$~&~0.5~&~$0.540\pm 0.067$~&~$0.538\pm 0.073$~&~$0.661\pm 0.065$~&
~$0.538\pm 0.070$~&~$0.537\pm 0.068$\\
~$~~$~&~$~~$~&~1.0~&~$0.544\pm 0.066$~&~$0.542\pm 0.074$~&~$0.924\pm 0.074$~&
~$0.533\pm 0.077$~&~$0.533\pm 0.070$\\
\hline
\end{tabular}
\caption{The effect of differential gain, pointing, beamwidth, ellipticity and rotation on parameter estimation for POLARBEAR obtained with MCMC simulations. 
The systematic beam parameters $\varepsilon$, $g$, etc. were chosen so that $\delta^{Fish}$ (Eq.26) assumes the specified values (third column from left), i.e. the bias 
-to-uncertainty ratio in the tensor-to-scalar ratio $r$ (assuming $r=$0.01),
as obtained by the Fisher-matrix-based calculation.
The values shown are the cosmological parameters recovered from the full likelihood 
function and their $1\sigma$ errors. The biases we obtain for differential beamwidth and ellipticity are orders of magnitude larger and are not shown here.} 
\end{table*}

\begin{table*}[c]
\begin{tabular}{|c|c|c|c|c|c|c|c|}
\hline
~Parameter~&~{\rm no sys.}~&~$\delta^{Fish}$~&~{\rm diff. gain}~&~{\rm diff. pointing}~&~{\rm diff. beamwidth}~&
~{\rm diff. ellipticity}~&~{\rm diff. rotation}~\\
\hline
~$r$~&~$0.00961\pm 0.00039$~&~0.1~&~$0.00964\pm 0.0040$~&~$0.00963\pm 0.0040$~&~$0.00964\pm 0.0039$~
&~$0.00962\pm 0.0039$~&~$0.00963\pm 0.0040$\\
~$$~&~$$~&~0.5~&~$0.00982\pm 0.0040$~&~$0.00979\pm 0.00041$~&~$0.00983\pm 0.00040$~
&~$0.00980\pm 0.00041$~&~$0.00980\pm 0.00040$\\
~$$~&~$$~&~1.0~&~$0.01004\pm 0.00041$~&~$0.01000\pm 0.00042$~&~$0.01004\pm 0.00041$~
&~$0.01002\pm 0.00040$~&~$0.00990\pm 0.00040$\\
\hline
~$w$~&~$-1.025\pm 0.097$~&~0.1~&~$-1.027\pm 0.095$~&~$-1.027\pm 0.097$~&~$-1.045\pm 0.100$~&
~$-1.025\pm 0.095$~&~$-1.029\pm 0.097$\\
~$$~&~$$~&~0.5~&-$1.027\pm 0.095$~&~$-1.028\pm 0.094$~&~$-1.152\pm 0.121$~&~$-1.021\pm 0.091$~&~$-1.032\pm 0.098$\\
~$$~&~$$~&~1.0~&-$1.026\pm 0.094$~&~$-1.027\pm 0.097$~&~$-1.304\pm 0.149$~&~$-1.026\pm 0.096$~&~$-1.026\pm 0.093$\\
\hline
~$M_{\nu}[{\rm eV}]$~&~$0.534\pm 0.014$~&~0.1~&~$0.534\pm 0.014$~&~$0.534\pm 0.014$~&~$0.538\pm 0.014$~&
~$0.534\pm 0.014$~&~$0.535\pm 0.014$\\
~$~~$~&~$~~$~&~0.5~&~$0.535\pm 0.014$~&~$0.535\pm 0.014$~&~$0.556\pm 0.013$~&
~$0.534\pm 0.014$~&~$0.535\pm 0.014$\\
~$~~$~&~$~~$~&~1.0~&~$0.535\pm 0.014$~&~$0.536\pm 0.014$~&~$0.577\pm 0.013$~&
~$0.534\pm 0.014$~&~$0.535\pm 0.014$\\
\hline
\end{tabular}
\caption{The effect of beam systematics on parameter estimation from CMBPOL-B obtained with using MCMC simulations.}
\end{table*}

\begin{table*}[c]
\begin{tabular}{|c|c|c|c|c|c|c|c|}
\hline
~Parameter~&~{\rm no sys.}~&~$\delta^{Fish}$~&~{\rm diff. gain}~&~{\rm diff. pointing}~&~{\rm diff. beamwidth}~&
~{\rm diff. ellipticity}~&~{\rm diff. rotation}~\\
\hline
~$r$~&~$0.01035\pm 0.00333$~&~0.1~&~$0.01076\pm 0.03221$~&~$0.01076\pm 0.00330$~&~$0.01084\pm 0.00323$~
&~$0.01053\pm 0.00342$~&~$0.01046\pm 0.00301$\\
~$$~&~$$~&~0.5~&~$0.01199\pm 0.00335$~&~$0.01241\pm 0.00365$~&~$0.01287\pm 0.00384$~
&~$0.01289\pm 0.00361$~&~$0.01226\pm 0.00361$\\
~$$~&~$$~&~1.0~&~$0.01445\pm 0.00358$~&~$0.01506\pm 0.00409$~&~$0.02326\pm 0.00639$~
&~$0.02135\pm 0.00587$~&~$0.01448\pm 0.00386$\\
\hline
~$w$~&~$-1.143\pm 0.362$~&~0.1~&~$-1.163\pm 0.349$~&~$-1.141\pm 0.352$~&~$-1.167\pm 0.361$~&
~$-1.131\pm 0.370$~&~$-1.143\pm 0.368$\\
~$$~&~$$~&~0.5~&-$1.123\pm 0.361$~&~$-1.109\pm 0.363$~&~$-1.228\pm 0.326$~&~$-1.131\pm 0.362$~&~$-1.121\pm 0.357$\\
~$$~&~$$~&~1.0~&-$1.137\pm 0.356$~&~$-1.153\pm 0.362$~&~$-1.347\pm 0.322$~&~$-1.160\pm 0.364$~&~$-1.137\pm 0.362$\\
\hline
~$M_{\nu}[{\rm eV}]$~&~$0.535\pm 0.109$~&~0.1~&~$0.536\pm 0.105$~&~$0.531\pm 0.114$~&~$0.548\pm 0.111$~&
~$0.522\pm 0.112$~&~$0.532\pm 0.110$\\
~$~~$~&~$~~$~&~0.5~&~$0.526\pm 0.120$~&~$0.526\pm 0.114$~&~$0.611\pm 0.110$~&
~$0.533\pm 0.124$~&~$0.530\pm 0.112$\\
~$~~$~&~$~~$~&~1.0~&~$0.528\pm 0.117$~&~$0.552\pm 0.114$~&~$0.774\pm 0.120$~&
~$0.530\pm 0.123$~&~$0.532\pm 0.112$\\
\hline
\end{tabular}
\caption{The effect of beam systematics on parameter estimation from QUIET+CLOVER obtained with using MCMC simulations.}
\end{table*}
\newpage

\begin{table*}[c]
\begin{tabular}{|c|c|c|c|c|}
\hline
~Experiment~&~{\rm Beam parameter}~&~{\rm $\delta_{r}$ ($\delta_{r}^{Fish}=0.1$)}~
&~{\rm $\delta_{r}$ ($\delta_{r}^{Fish}=0.5$)}~&~{\rm $\delta_{r}$ ($\delta_{r}^{Fish}=1.0$)}~\\
\hline
~{POLARBEAR}~&~$g\sqrt{\frac{f_{1}}{2\pi}}$~&~0.14~&~0.56~&~1.36~\\
~~&~$\rho\sqrt{\frac{f_{2}}{2\pi}}$~&~0.14~&~0.58~&~1.36~\\
~~&~$\mu\sqrt{\frac{f_{1}}{2\pi}}$~&~0.17~&~0.83~&~2.83~\\
~~&~$e$~&~0.11~&~0.64~&~2.47~\\
~~&~$\varepsilon$~&~0.17~&~0.53~&~1.23~\\
\hline
~{CMBPOL-B}~&~$g\sqrt{\frac{f_{1}}{2\pi}}$~&~0.08~&~0.54~&~1.10~\\
~~&~$\rho\sqrt{\frac{f_{2}}{2\pi}}$~&~0.05~&~0.46~&~1.15~\\
~~&~$\mu\sqrt{\frac{f_{1}}{2\pi}}$~&~0.08~&~0.56~&~1.10~\\
~~&~$e$~&~0.03~&~0.49~&~1.05~\\
~~&~$\varepsilon$~&~0.05~&~0.49~&~0.74~\\
\hline
~{QUIET+CLOVER}~&~$g\sqrt{\frac{f_{1}}{2\pi}}$~&~0.12~&~0.49~&~1.23~\\
~~&~$\rho\sqrt{\frac{f_{2}}{2\pi}}$~&~0.12~&~0.62~&~1.41~\\
~~&~$\mu\sqrt{\frac{f_{1}}{2\pi}}$~&~0.15~&~0.76~&~3.88~\\
~~&~$e$~&~0.06~&~0.76~&~3.30~\\
~~&~$\varepsilon$~&~0.04~&~0.57~&~1.24~\\
\hline
\end{tabular}
\caption{The bias in the tensor-to-scalar ratio $r$ ($\delta_{r}$) obtained with MCMC for POLARBEAR, CMBPOL-B and QUIET+CLOVER. These $\delta_{r}$ values were obtained by assuming each of the five systematics we considered have the values which induce a bias $\delta_{r}^{Fish}=0.1$, $0.5$ and $1$, respectively, in the Fisher matrix analysis. This table is a compilation of the corresponding values for $r$ reported in Tables X, XI and XII. Note the discrepancy between the Fisher-matrix-based and MCMC forecast for the bias for the differential beamwidth and ellipticity systematics. Both scale as the second-order gradient of the temperature $C_{l}^{sys}\propto l^{4}C_{l}^{T}$ and as a result of this steep rise of the systematics with scale the systematics soon overwhelm the primordial B-mode signal and significantly bias the deduced tensor-to-scalar ratio. The Fisher matrix estimate of the bias is only a leading order approximation in case the bias is small; an assumption which evidently does not apply to systematics which scale as the second order gradient of the temperature anisotropy.} 
\end{table*}

\end{document}